\documentstyle[11pt]{article}

\newcommand{\al}{\alpha}
\newcommand{\be}{\beta}
\newcommand{\ga}{\gamma}
\newcommand{\de}{\delta}
\newcommand{\ep}{\varepsilon}
\newcommand{\la}{\lambda}
\newcommand{\ro}{\rho}
\newcommand{\si}{\sigma}
\newcommand{\Tr}{\mathop{\rm Tr}\nolimits}
\newcommand{\dd}{\partial}
\newcommand{\q}{\overline{q}}
\newcommand{\G}{\widetilde{G}}
\newcommand{\va}[1]{\left<#1\right>}
\newcommand{\qq}[1]{\va{\q#1q}}
\newcommand{\as}{\al_s}
\newcommand{\D}{\overleftarrow{D}}
\newcommand{\p}{\hat{p}}
\newcommand{\Dp}{\left(\frac{dp}{2\pi}\right)_D}
\newcommand{\MS}{$\overline{\rm MS}$}
\newcommand{\PLB}{Phys.\ Lett.\ B}
\newcommand{\NPB}{Nucl.\ Phys.\ B}
\newcommand{\vv}{\leftrightarrow}

\begin{document}

\begin{flushright}
OUT--4102--54\\
hep-ph/9412238\\
December 1994
\end{flushright}

\vspace{2cm}

\begin{center}
\Large Methods of calculation\\
\Large of higher power corrections in QCD
\end{center}

\vspace{1cm}

\begin{center}
A.~G.~Grozin$^1$\\
Physics Department, Open University,\\
Milton Keynes MK7 6AA, UK
\end{center}

\vspace{1cm}

\begin{center}
Abstract
\end{center}

Although the methods of calculation of power corrections in QCD sum rules
are well known, algebraic complexity rapidly grows with the increase of
vacuum condensates' dimensions. Currently, state--of--the--art calculations
include dimension 7 and 8 condensates. I summarize and extend algorithms of
such calculations. First, I present all the formulae necessary for application
of the systematic classification of bilinear quark condensates proposed
earlier, and extend this method to the case of gluon condensates. Then I
apply these systematic procedures to expansions of bilinear and noncollinear
quark and gluon condensates in local ones, and of noncollinear condensates
in bilocal ones. The formulae obtained can be used for calculation of
correlators involving nonlocal condensates, and for inventing consistent
anzatzs for these condensates. Finally, I briefly summarize the methods of
calculation of heavy-- and light--quark currents' correlators. This paper
is aimed both to present new results on gluon and nonlocal condensates and
to be a self--contained handbook of formulae necessary for calculation of
power corrections in QCD sum rules.

\vfill
\footnoterule
\noindent
$^1$ A.Grozin@open.ac.uk; on leave of absence from
Budker Institute of Nuclear Physics, Novosibirsk 630090, Russia

\newpage

\section{Introduction}
\label{Intro}

The Operator Product Expansion (OPE) is widely used for investigation
of correlators in quantum field theory. In particular, it is
the basis of the QCD sum rules method. Let $\Pi(p)$ be a correlator
of two local currents in the momentum space. Then OPE reads
\begin{equation}
\Pi(p)=\sum_i c_i(p^2)\va{O_i},
\label{OPE}
\end{equation}
where $\va{O_i}$ are vacuum averages of local operators of various dimensions
$d$ (called also vacuum condensates), and $c_i(p^2)$ are coefficient functions.
They are perturbative series in $\as$. In order to make sum rules more precise
and to establish their applicability regions it is necessary to calculate both
higher perturbative corrections (higher terms in $\as$ expansions
of coefficient functions) and higher non--perturbative (power) corrections.
I shall not discuss physical applications of QCD sum rules here.

Methods of calculation of higher loop diagrams (perturbative corrections)
form an established field of research, and many articles and reviews are
devoted to them. I shall not discuss this subject; instead
I shall concentrate on the methods of calculation of higher--dimensional
terms in the OPE~(\ref{OPE}) in the leading order in $\as$.

There already exists an excellent review on technical problems
of QCD sum rules calculations~\cite{TechRev}. The present article can be
considered as an addendum to it. In all cases when a question is discussed
in~\cite{TechRev}, I shall refer to this review and the literature cited
in it. I should like to apologize to the authors of these articles
for not citing them directly.

I summarize and extend the methods of calculation of higher power corrections
to correlators of heavy--~\cite{NikRad1,NikRad} and the light--quark
currents~[4--9]. 
Full technical details (partly omitted from the original papers) will be
included for the reader's convenience. In many cases, even if an idea of
a calculation technique have appeared in the literature, it is very
tedious and error--prone to rederive all the necessary formulae for each
specific application. A set of reliable formulae can provide an invaluable
help for many calculations of power corrections to QCD sum rules.

I shall formulate the algorithms of calculation of higher power corrections
in a form suitable for a Computer Algebra implementation.
Unfortunately, there is no complete package for such calculations
in conventional Computer Algebra systems (REDUCE, Mathematica,\dots).
Such a package would be very useful. I have written several REDUCE
procedures and Modula-2 programs (producing a REDUCE readable output) which
implement some of the discussed algorithms. They are useful as a set of
tools though they can't be a substitute for a complete package.

The plan of the article is following. In Section~\ref{SecFPG} I briefly
introduce the fixed--point (Fock--Schwinger) gauge on which the modern
methods of calculation of power corrections are based; for more details
and proofs see~\cite{TechRev}. Section~\ref{SecCond} is devoted to
the systematic classification of vacuum condensates. There exist many
relations between condensates, therefore one has to choose a basis of
independent condensates and formulate an algorithm of reducing an arbitrary
condensate to this basis. A systematic classification of bilinear quark
condensates has been proposed in~\cite{GrPin2}. A carefully checked set
of formulae implementing this method is presented in Section~\ref{SubQuark};
this makes its application to any particular case very easy. In Section~%
\ref{SubGluon}, I extend this method to the case of gluon condensates.
Section~\ref{SubAver} contains some useful techniques of averaging two--
and three--point correlators.

Any calculation of power corrections to a correlator naturally involves
nonlocal condensates. In the standard approach, they are expanded in
series in local ones. Section~\ref{SecNonl} is devoted to a detailed
study of some simple types of nonlocal condensates using the systematic
methods introduced in Section~\ref{SecCond}. Tree--level calculations
of two--point correlators usually involve gauge--invariant bilocal
condensates. Their expansions in local ones are considered in Section~%
\ref{SubBil}. Three--point correlators, as well as correlators involving
bilocal currents (which are used in sum rules for hadron wave functions)
require more complicated trilocal objects---noncollinear condensates
(Section~\ref{SubNonc}). In some applications, such as sum rules for wave
function moments, it is convenient to expand noncollinear condensates
in series in one of the spacings. Coefficients in these series are bilocal
condensates. Such expansions are constructed in Section~\ref{SubColl}.
Expansions obtained in this Section can be used for calculation of
correlators involving nonlocal condensates, and for inventing consistent
anzatzs for these condensates.

Finally, I briefly summarize methods of calculation of heavy quark
currents' correlators (including heavy quark condensates) in Section~%
\ref{SecHeavy}, and of light quark currents' correlators--- in Section~%
\ref{SecLight}.

\section{Fixed--point gauge}
\label{SecFPG}\setcounter{equation}{0}

The QCD vacuum has a non--trivial structure. In order to calculate
a correlator in QCD, we should first calculate it in a background (vacuum)
gluon and quark field. Then we should average the expression for
a correlator via these fields over the vacuum. After that we obtain
the expression for a correlator via vacuum condensates~(\ref{OPE}).

Correlators of colourless currents are gauge invariant. Therefore we can use
any gauge for the background gluon field. It is convenient to use
the fixed--point (Fock--Schwinger) gauge
\begin{equation}
x_\mu A^a_\mu(x)=0.
\label{FPG}
\end{equation}
In this gauge the Taylor expansions for $A^a_\mu(x)$ and $q(x)$ can be
written in a gauge--covariant form~\cite{TechRev}
\begin{eqnarray}
&&A^a_\mu(x)=\frac1{2\cdot0!}x_\nu G^a_{\nu\mu}(0)
+\frac1{3\cdot1!}x_\al x_\nu D_\al G^a_{\nu\mu}(0)
\nonumber\\
&&\quad{}+\frac1{4\cdot2!}x_\be x_\al x_\nu D_\be D_\al G^a_{\nu\mu}(0)
+O(x^4),
\label{FPGexp}\\
&&q(x)=q(0)+x_\al D_\al q(0)+\frac1{2!}x_\be x_\al D_\be D_\al q(0)
+O(x^3)
\nonumber
\end{eqnarray}
(and hence $\q(x)=\q(0)+\q(0)\D_\al x_\al+\cdots$). Here
$D_\mu q=(\dd_\mu-iA_\mu)q$ is a covariant derivative in the fundamental
representation, $A_\mu=gA^a_\mu t^a$;
$D_\mu G^a_{\ro\si}=(\dd_\mu\de^{ab}+A^{ab}_\mu)G^b_{\ro\si}$
is a covariant derivative in the adjoint representation,
$A^{ab}_\mu=gf^{acb}A^c_\mu$. Hence only gauge covariant quantities
are used at all stages of calculations.

In this gauge the theory is not translational invariant. A translation
should be accompanied by a gauge transformation to another fixed point gauge.
Correlators become gauge invariant and hence translationally invariant after
vacuum averaging. We can choose the gauge fixed point (origin in~(\ref{FPG}))
at any vertex of the correlator to simplify calculations. In the momentum
space, background gluon lines have incoming momenta $k_i$ and contribute
$A^a_\mu(k_i)\frac{d^4k_i}{(2\pi)^4}$ where $A^a_\mu(k_i)$ is the Fourier
transform of~(\ref{FPGexp}) and is the series in derivatives of $\de(k_i)$.
The vertex chosen as the gauge fixed point provides a common sink for all
vacuum momenta $k_i$.

\input FEYNMAN

\begin{figure}[ht]
\begin{picture}(40000,22000)
\global\Yone=15000 \global\Xone=300
\THICKLINES
\drawline\fermion[\W\REG](10000,\Yone)[8000]
\drawarrow[\W\ATBASE](\pmidx,\pmidy)
\THINLINES
\global\Xtwo=\pmidx \global\Ytwo=\pmidy
\global\Xthree=\pbackx \global\Ythree=\pbacky
\global\Xfour=\pfrontx \global\Yfour=\pfronty
\global\advance\pmidy by -1000
\global\advance\pmidx by 1500
\drawline\fermion[\W\REG](\pmidx,\pmidy)[3000]
\drawarrow[\W\ATTIP](\pbackx,\pbacky)
\global\advance\Ytwo by -2000
\put(\Xtwo,\Ytwo){\makebox(0,0)[t]{$p$}}
\global\advance\Ytwo by 3000
\put(\Xtwo,\Ytwo){\makebox(0,0)[b]{$iS(p)$}}
\put(\Xthree,\Ythree){\circle*{\Xone}}
\global\advance\Ythree by -1000
\put(\Xthree,\Ythree){\makebox(0,0)[t]{$x$}}
\put(\Xfour,\Yfour){\circle*{\Xone}}
\global\advance\Yfour by -1000
\put(\Xfour,\Yfour){\makebox(0,0)[t]{$0$}}
\global\advance\Yfour by 1000
\global\advance\Xfour by 1000
\put(\Xfour,\Yfour){\makebox(0,0)[l]{$=$}}
\drawline\fermion[\W\REG](21000,\Yone)[8000]
\drawarrow[\W\ATBASE](\pmidx,\pmidy)
\global\Xtwo=\pmidx \global\Ytwo=\pmidy
\put(\pbackx,\pbacky){\circle*{\Xone}}
\put(\pfrontx,\pfronty){\circle*{\Xone}}
\global\advance\pfrontx by 1000
\put(\pfrontx,\pfronty){\makebox(0,0)[l]{$+$}}
\global\advance\pmidy by -1000
\global\advance\pmidx by 1500
\drawline\fermion[\W\REG](\pmidx,\pmidy)[3000]
\drawarrow[\W\ATTIP](\pbackx,\pbacky)
\global\advance\Ytwo by -2000
\put(\Xtwo,\Ytwo){\makebox(0,0)[t]{$p$}}
\global\advance\Ytwo by 3000
\put(\Xtwo,\Ytwo){\makebox(0,0)[b]{$iS_0(p)$}}
\drawline\fermion[\W\REG](30000,\Yone)[6000]
\drawarrow[\W\ATBASE](\pmidx,\pmidy)
\global\Xtwo=\pmidx \global\Ytwo=\pmidy
\put(\pbackx,\pbacky){\circle*{\Xone}}
\global\advance\Ytwo by -2000
\put(\Xtwo,\Ytwo){\makebox(0,0)[t]{$p$}}
\global\advance\pmidy by -1000
\global\advance\pmidx by 1500
\drawline\fermion[\W\REG](\pmidx,\pmidy)[3000]
\drawarrow[\W\ATTIP](\pbackx,\pbacky)
\drawline\fermion[\W\REG](36000,\Yone)[6000]
\drawarrow[\W\ATBASE](\pmidx,\pmidy)
\global\Xtwo=\pmidx \global\Ytwo=\pmidy
\put(\pfrontx,\pfronty){\circle*{\Xone}}
\global\advance\pfrontx by 1000
\put(\pfrontx,\pfronty){\makebox(0,0)[l]{$+$}}
\global\advance\Ytwo by -2000
\put(\Xtwo,\Ytwo){\makebox(0,0)[t]{$p-k_1$}}
\global\advance\pmidy by -1000
\global\advance\pmidx by 1500
\drawline\fermion[\W\REG](\pmidx,\pmidy)[3000]
\drawarrow[\W\ATTIP](\pbackx,\pbacky)
\drawline\photon[\N\REG](30000,\Yone)[6]
\global\advance\pmidx by 1000
\global\advance\pmidy by 1500
\drawline\fermion[\S\REG](\pmidx,\pmidy)[3000]
\drawarrow[\S\ATTIP](\pbackx,\pbacky)
\global\advance\pmidx by 1000
\put(\pmidx,\pmidy){\makebox(0,0)[l]{$k_1$}}
\global\Yone=5000
\drawline\fermion[\W\REG](8000,\Yone)[6000]
\drawarrow[\W\ATBASE](\pmidx,\pmidy)
\global\Xtwo=\pmidx \global\Ytwo=\pmidy
\put(\pbackx,\pbacky){\circle*{\Xone}}
\global\advance\Ytwo by -2000
\put(\Xtwo,\Ytwo){\makebox(0,0)[t]{$p$}}
\global\advance\pmidy by -1000
\global\advance\pmidx by 1500
\drawline\fermion[\W\REG](\pmidx,\pmidy)[3000]
\drawarrow[\W\ATTIP](\pbackx,\pbacky)
\drawline\fermion[\W\REG](14000,\Yone)[6000]
\drawarrow[\W\ATBASE](\pmidx,\pmidy)
\global\Xtwo=\pmidx \global\Ytwo=\pmidy
\global\advance\Ytwo by -2000
\put(\Xtwo,\Ytwo){\makebox(0,0)[t]{$p-k_1$}}
\global\advance\pmidy by -1000
\global\advance\pmidx by 1500
\drawline\fermion[\W\REG](\pmidx,\pmidy)[3000]
\drawarrow[\W\ATTIP](\pbackx,\pbacky)
\drawline\fermion[\W\REG](20000,\Yone)[6000]
\drawarrow[\W\ATBASE](\pmidx,\pmidy)
\put(\pfrontx,\pfronty){\circle*{\Xone}}
\global\advance\pfrontx by 1000
\put(\pfrontx,\pfronty){\makebox(0,0)[l]{$+\cdots$}}
\global\Xtwo=\pmidx \global\Ytwo=\pmidy
\global\advance\Ytwo by -2000
\put(\Xtwo,\Ytwo){\makebox(0,0)[t]{$p-k_1-k_2$}}
\global\advance\pmidy by -1000
\global\advance\pmidx by 1500
\drawline\fermion[\W\REG](\pmidx,\pmidy)[3000]
\drawarrow[\W\ATTIP](\pbackx,\pbacky)
\drawline\photon[\N\REG](8000,\Yone)[6]
\global\advance\pmidx by 1000
\global\advance\pmidy by 1500
\drawline\fermion[\S\REG](\pmidx,\pmidy)[3000]
\drawarrow[\S\ATTIP](\pbackx,\pbacky)
\global\advance\pmidx by 1000
\put(\pmidx,\pmidy){\makebox(0,0)[l]{$k_1$}}
\drawline\photon[\N\REG](14000,\Yone)[6]
\global\advance\pmidx by 1000
\global\advance\pmidy by 1500
\drawline\fermion[\S\REG](\pmidx,\pmidy)[3000]
\drawarrow[\S\ATTIP](\pbackx,\pbacky)
\global\advance\pmidx by 1000
\put(\pmidx,\pmidy){\makebox(0,0)[l]{$k_2$}}
\end{picture}
\caption{Quark propagator in background gluon field}
\label{FigQprop}
\end{figure}
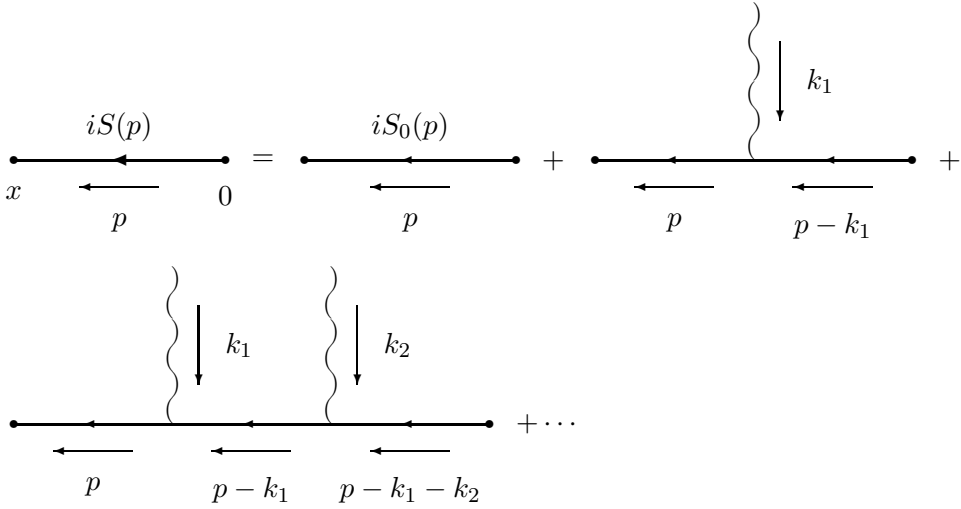

The quark propagator (Fig.~\ref{FigQprop}) can be written as a sum over
the number of background gluon lines:
\begin{eqnarray}
&&S(p)=\sum_{k=0}^\infty S_k(p), \quad
S_0(p)=\frac{\p+m}{p^2-m^2}, \quad
S_k(p)=-S_0(p)\hat A S_{k-1}(p),
\nonumber\\
&&A_\mu=-i\sum_{l=0}^\infty\frac{(-i)^l}{(l+2)l!}
D_{\al_1}\ldots D_{\al_l}G^a_{\nu\mu}t^a\dd_{\al_1}\ldots\dd_{\al_l}\dd_\nu.
\label{Sp}
\end{eqnarray}
It is convenient to use recurrent relations for $S_k(p)$:
\begin{equation}
S_k=S_0\sum_{l=0}^\infty\frac{A_{kl}}{l+2}, \quad
A_{k0}=iG_{\mu\nu}\dd_\mu\ga_\nu S_{k-1}, \quad
A_{kl}=-\frac{i}{l}\dd_\al D_\al A_{k,l-1}.
\label{Sk}
\end{equation}
Here $\dd_\mu=\dd/\dd p_\mu$ acts on all propagators to the right, and
$D_\mu$---only on the nearest $G_{\mu\nu}=gG^a_{\mu\nu}(0)t^a$. For example,
retaining gluon operators with $d\le4$ (Fig.~\ref{FigQprop}) we have
\begin{eqnarray}
&&S=S_0+\frac{i}2G_{\mu\nu}S_0\dd_\mu\ga_\nu S_0
+\frac13D_\al G_{\mu\nu} S_0\dd_\al\dd_\mu\ga_\nu S_0
\label{S4s}\\
&&\quad{}-\frac{i}8D_\be D_\al G_{\mu\nu} S_0\dd_\be\dd_\al\dd_\mu\ga_\nu S_0
-\frac14G_{\ro\si}G_{\mu\nu}S_0\dd_\ro\ga_\si S_0\dd_\mu\ga_\nu S_0.
\nonumber
\end{eqnarray}

I have written a Modula-2 program for simplification of expressions
constructed from $S_0$, $\dd_\mu$, and $\ga_\mu$. It applies $\dd_\mu$
to all $S_0$ to the right using $\dd_\mu S_0=-S_0\ga_\mu S_0$, collects
similar terms, and produces a REDUCE readable output, which can be used,
for example, to expand these expressions in basis $\ga$ matrices.

These formulae can be used for both massive and massless quarks.
For a light quark we can expand $S_0$ in $m$. For a massless quark
the equation~(\ref{S4s}) gives
\begin{eqnarray}
&&S(p)=\frac1{p^2}\p-\frac1{p^4}p_\mu\G_{\mu\nu}\ga_\nu\ga_5
\label{S4r}\\
&&\quad{}+\frac1{p^6}\Bigg[-\frac23\left(p^2\hat J-p_\mu J_\mu\p
+p_\la D_\la p_\mu G_{\mu\nu}\ga_\nu\right)
-2ip_\la D_\la p_\mu\G_{\mu\nu}\ga_\nu\ga_5\Bigg]
\nonumber\\
&&\quad{}+\frac1{p^8}\Bigg[2\left(p^2\ga_\mu G_{\mu\la}G_{\la\nu}p_\nu
-p_\mu G_{\mu\la}G_{\la\nu}p_\nu\p\right)
-2ip_\la D_\la\left(p^2\hat J-p_\mu J_\mu\p\right)
\nonumber\\
&&\quad{}+\left(4(p_\la D_\la)^2-p^2D^2\right)p_\mu\G_{\mu\nu}\ga_\nu\ga_5
-i\left[p_\mu G_{\mu\la},\G_{\la\nu}\ga_\nu\ga_5\right]\Bigg].
\nonumber
\end{eqnarray}
Here $J_\mu=gJ^a_\mu t^a$,
$J^a_\mu=D_\nu G^a_{\mu\nu}=g\sum_{q'}\q'\ga_\mu t^a q'$. Note that
the last term in~(\ref{S4r}) is missing in~\cite{TechRev}. In deriving this
formula, we have used the relations
$D^2G_{\mu\nu}=-D_\mu J_\nu+D_\nu J_\mu+2i[G_{\mu\la},G_{\la\nu}]$,
$D_\nu D_\la G_{\mu\nu}=D_\la J_\mu-i[G_{\nu\la},G_{\mu\nu}]$,
$G_{\mu\la}\G_{\la\nu}+\G_{\nu\la}G_{\la\mu}=-\frac12\de_{\mu\nu}
\G_{\ro\si}G_{\ro\si}$, $G_{\mu\nu}\G_{\mu\nu}=\G_{\mu\nu}G_{\mu\nu}$.
In practical calculations it is often more convenient to substitute
symbolic expressions for propagators like~(\ref{S4s}) to the diagrams
and to simplify the complete answers rather than to use~(\ref{S4r}).

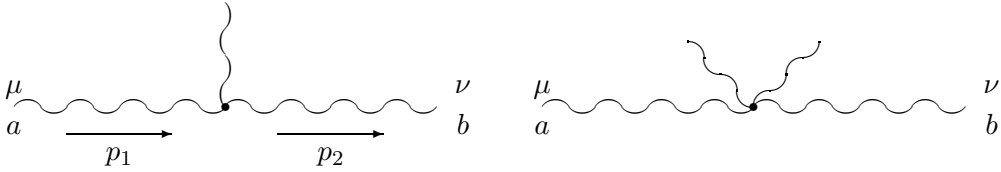
\begin{figure}[ht]
\begin{picture}(40000,7500)
\global\Yone=3000
\THINLINES
\put(10000,\Yone){\circle*{300}}
\drawline\photon[\W\REG](10000,\Yone)[8]
\drawline\photon[\E\REG](10000,\Yone)[8]
\drawline\photon[\N\REG](10000,\Yone)[4]
\put(30000,\Yone){\circle*{300}}
\drawline\photon[\W\REG](30000,\Yone)[8]
\drawline\photon[\E\REG](30000,\Yone)[8]
\drawline\photon[\NW\REG](30000,\Yone)[4]
\drawline\photon[\NE\REG](30000,\Yone)[4]
\global\advance\Yone by -1000
\drawline\fermion[\E\REG](4000,\Yone)[4000]
\drawarrow[\E\ATTIP](\pbackx,\pbacky)
\drawline\fermion[\E\REG](12000,\Yone)[4000]
\drawarrow[\E\ATTIP](\pbackx,\pbacky)
\put(2000,\Yone){\makebox(0,0)[b]{$a$}}
\put(19000,\Yone){\makebox(0,0)[b]{$b$}}
\put(22000,\Yone){\makebox(0,0)[b]{$a$}}
\put(39000,\Yone){\makebox(0,0)[b]{$b$}}
\global\advance\Yone by -1000
\put(6000,\Yone){\makebox(0,0)[b]{$p_1$}}
\put(14000,\Yone){\makebox(0,0)[b]{$p_2$}}
\global\advance\Yone by 3000
\put(2000,\Yone){\makebox(0,0)[t]{$\mu$}}
\put(19000,\Yone){\makebox(0,0)[t]{$\nu$}}
\put(22000,\Yone){\makebox(0,0)[t]{$\mu$}}
\put(39000,\Yone){\makebox(0,0)[t]{$\nu$}}
\end{picture}
\caption{Vertices of gluon's interaction with background gluon field}
\label{FigVert}
\end{figure}

In order to consider the gluon propagation in the background gluon field
we should substitute $A^a_\mu+a^a_\mu$ into the QCD lagrangian, where
$A^a_\mu$ is the background field and $a^a_\mu$ is the quantum gluon field.
The lagrangian becomes
\begin{equation}
L=-\frac14G^a_{\mu\nu}G^a_{\mu\nu}
-\frac12(D_\mu a^a_\nu)(D_\mu a^a_\nu)
+\frac12(D_\mu a^a_\nu)(D_\nu a^a_\mu)
+\frac12a^a_\mu G^{ab}_{\mu\nu}a^b_\nu+\cdots
\label{Lagr}
\end{equation}
Here $G^a_{\mu\nu}$ and $D_\mu$ include only the background field $A^a_\mu$,
and $G^{ab}_{\mu\nu}=gf^{acb}G^c_{\mu\nu}$. We choose the fixed--point gauge
for the background field $A^a_\mu$. This lagrangian is still gauge invariant
with respect to $a^a_\mu$. We should add a gauge fixing term and
a corresponding ghost term. It is convenient to use a generalization
of the Feynman gauge fixing term $-\frac12(D_\mu a^a_\mu)^2$. Then the free
gluon propagator is $D^0_{\mu\nu}(p)=\de_{\mu\nu}/p^2$, and the vertices
of the gluon's interaction with the background field (Fig.~\ref{FigVert}) are
$iA^{ab}_\la(p_1+p_2)_\la\de_{\mu\nu}-iA^{ab}_\mu(p_1-p_2)_\nu
-iA^{ab}_\nu(p_2-p_1)_\mu-iG^{ab}_{\mu\nu}$
and $-A^{ac}_\la A^{cb}_\la\de_{\mu\nu}-A^{ac}_\mu A^{cb}_\nu
+A^{ac}_\mu A^{cb}_\nu$.

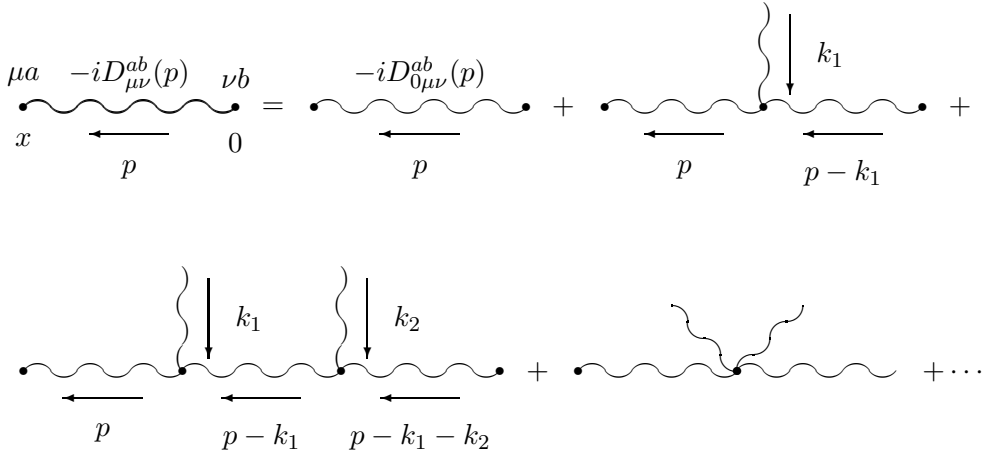
\begin{figure}[ht]
\begin{picture}(40000,22000)
\global\Yone=15000 \global\Xone=300
\THICKLINES
\drawline\photon[\W\REG](10000,\Yone)[8]
\THINLINES
\global\Xtwo=\pmidx \global\Ytwo=\pmidy
\global\Xthree=\pbackx \global\Ythree=\pbacky
\global\Xfour=\pfrontx \global\Yfour=\pfronty
\global\advance\pmidy by -1000
\global\advance\pmidx by 1500
\drawline\fermion[\W\REG](\pmidx,\pmidy)[3000]
\drawarrow[\W\ATTIP](\pbackx,\pbacky)
\global\advance\Ytwo by -2000
\put(\Xtwo,\Ytwo){\makebox(0,0)[t]{$p$}}
\global\advance\Ytwo by 3000
\put(\Xtwo,\Ytwo){\makebox(0,0)[b]{$-iD^{ab}_{\mu\nu}(p)$}}
\put(\Xthree,\Ythree){\circle*{\Xone}}
\global\advance\Ythree by -1000
\put(\Xthree,\Ythree){\makebox(0,0)[t]{$x$}}
\global\advance\Ythree by 2500
\put(\Xthree,\Ythree){\makebox(0,0)[t]{$\mu a$}}
\put(\Xfour,\Yfour){\circle*{\Xone}}
\global\advance\Yfour by -1000
\put(\Xfour,\Yfour){\makebox(0,0)[t]{$0$}}
\global\advance\Yfour by 2500
\put(\Xfour,\Yfour){\makebox(0,0)[t]{$\nu b$}}
\global\advance\Yfour by -1500
\global\advance\Xfour by 1000
\put(\Xfour,\Yfour){\makebox(0,0)[l]{$=$}}
\drawline\photon[\W\REG](21000,\Yone)[8]
\global\Xtwo=\pmidx \global\Ytwo=\pmidy
\put(\pbackx,\pbacky){\circle*{\Xone}}
\put(\pfrontx,\pfronty){\circle*{\Xone}}
\global\advance\pfrontx by 1000
\put(\pfrontx,\pfronty){\makebox(0,0)[l]{$+$}}
\global\advance\pmidy by -1000
\global\advance\pmidx by 1500
\drawline\fermion[\W\REG](\pmidx,\pmidy)[3000]
\drawarrow[\W\ATTIP](\pbackx,\pbacky)
\global\advance\Ytwo by -2000
\put(\Xtwo,\Ytwo){\makebox(0,0)[t]{$p$}}
\global\advance\Ytwo by 3000
\put(\Xtwo,\Ytwo){\makebox(0,0)[b]{$-iD^{ab}_{0\mu\nu}(p)$}}
\drawline\photon[\W\REG](30000,\Yone)[6]
\global\Xtwo=\pmidx \global\Ytwo=\pmidy
\put(\pbackx,\pbacky){\circle*{\Xone}}
\global\advance\Ytwo by -2000
\put(\Xtwo,\Ytwo){\makebox(0,0)[t]{$p$}}
\global\advance\pmidy by -1000
\global\advance\pmidx by 1500
\drawline\fermion[\W\REG](\pmidx,\pmidy)[3000]
\drawarrow[\W\ATTIP](\pbackx,\pbacky)
\drawline\photon[\E\REG](30000,\Yone)[6]
\global\Xtwo=\pmidx \global\Ytwo=\pmidy
\put(\pbackx,\pbacky){\circle*{\Xone}}
\global\advance\pbackx by 1000
\put(\pbackx,\pbacky){\makebox(0,0)[l]{$+$}}
\global\advance\Ytwo by -2000
\put(\Xtwo,\Ytwo){\makebox(0,0)[t]{$p-k_1$}}
\global\advance\pmidy by -1000
\global\advance\pmidx by 1500
\drawline\fermion[\W\REG](\pmidx,\pmidy)[3000]
\drawarrow[\W\ATTIP](\pbackx,\pbacky)
\drawline\photon[\N\REG](30000,\Yone)[4]
\put(30000,\Yone){\circle*{\Xone}}
\global\advance\pmidx by 1000
\global\advance\pmidy by 1500
\drawline\fermion[\S\REG](\pmidx,\pmidy)[3000]
\drawarrow[\S\ATTIP](\pbackx,\pbacky)
\global\advance\pmidx by 1000
\put(\pmidx,\pmidy){\makebox(0,0)[l]{$k_1$}}
\global\Yone=5000
\drawline\photon[\W\REG](8000,\Yone)[6]
\global\Xtwo=\pmidx \global\Ytwo=\pmidy
\put(\pbackx,\pbacky){\circle*{\Xone}}
\global\advance\Ytwo by -2000
\put(\Xtwo,\Ytwo){\makebox(0,0)[t]{$p$}}
\global\advance\pmidy by -1000
\global\advance\pmidx by 1500
\drawline\fermion[\W\REG](\pmidx,\pmidy)[3000]
\drawarrow[\W\ATTIP](\pbackx,\pbacky)
\drawline\photon[\E\REG](8000,\Yone)[6]
\global\Xtwo=\pmidx \global\Ytwo=\pmidy
\global\Xthree=\pbackx
\global\advance\Ytwo by -2000
\put(\Xtwo,\Ytwo){\makebox(0,0)[t]{$p-k_1$}}
\global\advance\pmidy by -1000
\global\advance\pmidx by 1500
\drawline\fermion[\W\REG](\pmidx,\pmidy)[3000]
\drawarrow[\W\ATTIP](\pbackx,\pbacky)
\drawline\photon[\E\REG](\Xthree,\Yone)[6]
\put(\pbackx,\pbacky){\circle*{\Xone}}
\global\advance\pbackx by 1000
\put(\pbackx,\pbacky){\makebox(0,0)[l]{$+$}}
\global\Xtwo=\pmidx \global\Ytwo=\pmidy
\global\advance\Ytwo by -2000
\put(\Xtwo,\Ytwo){\makebox(0,0)[t]{$p-k_1-k_2$}}
\global\advance\pmidy by -1000
\global\advance\pmidx by 1500
\drawline\fermion[\W\REG](\pmidx,\pmidy)[3000]
\drawarrow[\W\ATTIP](\pbackx,\pbacky)
\drawline\photon[\N\REG](8000,\Yone)[4]
\put(8000,\Yone){\circle*{\Xone}}
\global\advance\pmidx by 1000
\global\advance\pmidy by 1500
\drawline\fermion[\S\REG](\pmidx,\pmidy)[3000]
\drawarrow[\S\ATTIP](\pbackx,\pbacky)
\global\advance\pmidx by 1000
\put(\pmidx,\pmidy){\makebox(0,0)[l]{$k_1$}}
\drawline\photon[\N\REG](\Xthree,\Yone)[4]
\put(\Xthree,\Yone){\circle*{\Xone}}
\global\advance\pmidx by 1000
\global\advance\pmidy by 1500
\drawline\fermion[\S\REG](\pmidx,\pmidy)[3000]
\drawarrow[\S\ATTIP](\pbackx,\pbacky)
\global\advance\pmidx by 1000
\put(\pmidx,\pmidy){\makebox(0,0)[l]{$k_2$}}
\drawline\photon[\W\REG](29000,\Yone)[6]
\put(\pbackx,\pbacky){\circle*{\Xone}}
\drawline\photon[\E\REG](29000,\Yone)[6]
\put(\pfrontx,\pfronty){\circle*{\Xone}}
\global\advance\pbackx by 1000
\put(\pbackx,\pbacky){\makebox(0,0)[l]{$+\cdots$}}
\put(29000,\Yone){\circle*{\Xone}}
\drawline\photon[\NW\REG](29000,\Yone)[4]
\drawline\photon[\NE\REG](29000,\Yone)[4]
\end{picture}
\caption{Gluon propagator in background gluon field}
\label{FigGprop}
\end{figure}

Retaining terms with $d\le4$, we have the gluon propagator
(Fig.~\ref{FigGprop})
\begin{eqnarray}
&&D_{\al\be}=\frac1{p^2}\de_{\al\be}+\frac1{p^4}2G_{\al\be}
+\frac1{p^6}\left(\frac23ip_\mu J_\mu\de_{\al\be}
+4ip_\la D_\la G_{\al\be}\right)
\nonumber\\
&&\quad{}+\frac1{p^8}\Bigg[-2p_\la D_\la p_\mu J_\mu\de_{\al\be}
-2\left(4(p_\la D_\la)^2-p^2D^2\right)G_{\al\be}
\label{D4}\\
&&\quad{}+\frac12\left(p^2G_{\mu\nu}G_{\mu\nu}
+4p_\mu G_{\mu\la}G_{\la\nu}p_\nu\right)\de_{\al\be}
+4p^2G_{\al\la}G_{\la\be}\Bigg].
\nonumber
\end{eqnarray}
Here the matrix notations are used for colour indices.

\section{Vacuum condensates' classification}
\label{SecCond}\setcounter{equation}{0}

Vacuum condensates can be divided into classes according to the number
of quark fields in them. Those without quark fields are gluon condensates.
Those with two quark fields are bilinear quark condensates; they have $d\ge3$.
Four--quark condensates have $d\ge6$, and so on. The unit operator is
the gluon operator with $d=0$, according to this classification.

\begin{sloppypar}
As is clear from the previous Section, bilinear quark condensates of the form
$\qq{(D\ldots DG)\allowbreak(D\ldots DG)\ldots\allowbreak\ga\ldots\ga}$
appear in calculations
of bilinear quark currents' correlators at the tree level. Analogously,
gluon condensates of the form $\va{\Tr(D\ldots DG)(D\ldots DG)\ldots}$
appear in the one--loop approximation (there are at least two $(D\ldots DG)$
groups because otherwise the colour trace vanishes).
\end{sloppypar}

In Section~\ref{SubQuark} I discuss the systematic classification
of bilinear quark condensates following~\cite{GrPin2}. All formulae necessary
for the practical use of this method are presented. In Section~\ref{SubGluon}
I apply similar methods to the gluon condensates. Techniques of vacuum
averaging of expressions for correlators via quark and gluon fields are
discussed in Section~\ref{SubAver}.

\subsection{Bilinear quark condensates}
\label{SubQuark}

Using the formulae $G_{\mu\nu}=gG^a_{\mu\nu}t^a=i[D_\mu,D_\nu]$,
$(D_\mu A)=(D^{ab}_\mu A^b)t^a=[D_\mu,A]$ (where $A^a$ is in the adjoint
representation of the colour group, and $A=A^a t^a$), we can easily reduce
any bilinear quark condensate of dimension $d=m+3$ to a combination of terms
of the form $\qq{D_{\mu_1}\ldots D_{\mu_m}\Gamma_{\mu_1\ldots\mu_m}}$,
where $\Gamma_{\mu_1\ldots\mu_m}$ is constructed from $\ga_\mu$ and
$\de_{\mu\nu}$. Choosing the terms with the largest number
of $\ga$ matrices, we permute them in such a way that their indices are
in the same order as in $D_{\mu_1}\ldots D_{\mu_m}$. Arising additional
terms have fewer $\ga$ matrices. We repeat this procedure going to terms
with fewer $\ga$ matrices until we reach terms with one $\ga$ matrix or
without them at all. As a result, a bilinear quark condensate reduces
to a linear combination of terms of the form $\qq{O_i}$, where $O_i$
are constructed from $\hat D$ and $D_\mu$. Due to the equations of motion,
those terms in which $\hat D$ is adjacent to $q$ or $\q$ reduce to
lower--dimensional condensates multiplied by $m$.

Having written down all the condensates $B^d_i=\qq{O^m_i}$ where $O^m_i$ are
constructed from $\hat D$ and $D_\mu$ and no $\hat D$ is adjacent to $q$
and $\q$, we obtain a certain set of $d$--dimensional bilinear quark
condensates. We have just demonstrated that any $d$--dimensional condensate
can be systematically expressed via $B^d_i$, $mB^{d-1}_i$,\dots\ This means
that they form a basis of quark condensates with dimensions $\le d$. But this
basis is extremely inconvenient for use because its condensates contain
a maximum number of derivatives acting on $q$. So we should better choose
a basis of the most convenient $d$--dimensional condensates $Q^d_j$. We can
express $Q^d_j$ via $B^d_i$, $mB^{d-1}_i$,\dots\ Solving this linear system,
we obtain the expressions for $B^d_i$ via $Q^d_j$, $mQ^{d-1}_j$,\dots\
Having these expressions, we can easily reduce any $d$--dimensional bilinear
quark condensate to the basis ones $Q^d_j$, $mQ^{d-1}_j$,\dots

A general guideline for choosing good $Q^d_j$ is to have a minimum number
of derivatives acting on the quark field, and hence a maximum dimension
of gluon operators in $Q^d_j$. It allows to expand propagators to a minimum
dimension in the background gluon field. Among such operators one should
first of all choose those containing $J_\mu$ because they are really
four--quark ones and are more easily calculable. They may be suppressed
in some vacuum models (of the instanton type) in which vacuum gluon fields
may be strong but $D_\nu G_{\mu\nu}\approx0$.

For dimensions $d\le6$ we have
\begin{equation}
B^3=\qq{}, \quad
B^5=-\qq{D^2}, \quad
B^6=-i\qq{D_\mu\hat DD_\mu}.
\label{B6}
\end{equation}
We choose the basis condensates
\begin{equation}
Q^3=\qq{}, \quad
Q^5=i\qq{G_{\mu\nu}\si_{\mu\nu}}, \quad
Q^6=\qq{\hat J},
\label{Q6}
\end{equation}
where $\si_{\mu\nu}=\ga_{[\mu}\ga_{\nu]}$, and square brackets mean
antisymmetrization ($Q^5$ is usually denoted $m_0^2Q^3$).
These condensates are expressed via $B^d_i$ as
\begin{eqnarray}
&&Q^3=B^3,
\nonumber\\
&&Q^5=-2B^5+2m^2B^3,
\label{QB6}\\
&&Q^6=-2B^6+2mB^5.
\nonumber
\end{eqnarray}
Solving this linear system we obtain
\begin{eqnarray}
&&B^3=Q^3,
\nonumber\\
&&B^5=-\frac12Q^5+m^2Q^3,
\label{BQ6}\\
&&B^6=-\frac12Q^6-\frac{m}2Q^5+m^3Q^3.
\nonumber
\end{eqnarray}

For $d=7$ we have
\begin{eqnarray}
&&B^7_1=\qq{D^2D^2}, \quad
B^7_2=\qq{D_\mu D^2D_\mu},
\nonumber\\
&&B^7_3=\qq{D_\mu D_\nu D_\mu D_\nu}, \quad
B^7_4=\qq{D_\mu \hat D^2D_\mu}.
\label{B7}
\end{eqnarray}
We choose basis condensates
\begin{eqnarray}
&&Q^7_1=\qq{G_{\mu\nu}G_{\mu\nu}}, \quad
Q^7_2=i\qq{G_{\mu\nu}\G_{\mu\nu}\ga_5},
\nonumber\\
&&Q^7_3=\qq{G_{\mu\la}G_{\la\nu}\si_{\mu\nu}}, \quad
Q^7_4=i\qq{D_\mu J_\nu\si_{\mu\nu}}.
\label{Q7}
\end{eqnarray}
Here the condensate containing
$\G_{\mu\nu}=\frac12\ep_{\mu\nu\ro\si}G_{\ro\si}$ and
$\ga_5=\frac{i}{4!}\ep_{\al\be\ga\de}\ga_\al\ga_\be\ga_\ga\ga_\de$
is understood as a short notation for the expression from which both $\ep$
tensors are eliminated using $\ep^{\mu\nu\ro\si}\ep_{\al\be\ga\de}
=-4!\de^\mu_{[\al}\de^\nu_\be\de^\ro_\ga\de^\si_{\de]}$,
and this expression is valid at any space dimension $D$. These condensates
are expressed via $B^d_i$ as
\begin{eqnarray}
&&Q^7_1=2B^7_2-2B^7_3,
\nonumber\\
&&Q^7_2=-2B^7_1-2B^7_2+2B^7_3+2B^7_4-4mB^6+6m^2B^5-2m^4B^3,
\nonumber\\
&&Q^7_3=2B^7_2-2B^7_3-B^7_4+2mB^6-m^2B^5,
\label{QB7}\\
&&Q^7_4=-2B^7_1-2B^7_2+4B^7_3-4mB^6+4m^2B^5.
\nonumber
\end{eqnarray}
Solving this linear system we obtain
\begin{eqnarray}
&&B^7_1=\frac12Q^7_1-\frac12Q^7_2-Q^7_3-m^2Q^5+m^4Q^3,
\nonumber\\
&&B^7_2=\frac32Q^7_1-\frac12Q^7_2-Q^7_3+\frac12Q^7_4-mQ^6-m^2Q^5+m^4Q^3,
\nonumber\\
&&B^7_3=Q^7_1-\frac12Q^7_2-Q^7_3+\frac12Q^7_4-mQ^6-m^2Q^5+m^4Q^3,
\label{BQ7}\\
&&B^7_4=Q^7_1-Q^7_3-mQ^6-\frac{m^2}2Q^5+m^4Q^3.
\nonumber
\end{eqnarray}

Similarly, for $d=8$ we have
\begin{eqnarray}
&&B^8_1=i\qq{(D_\mu\hat DD_\mu D^2+D^2D_\mu\hat DD_\mu)},
\nonumber\\
&&B^8_2=i\qq{(D_\mu\hat DD^2D_\mu+D_\mu D^2\hat DD_\mu)},
\nonumber\\
&&B^8_3=i\qq{(D_\mu\hat DD_\nu D_\mu D_\nu+D_\nu D_\mu D_\nu\hat DD_\mu)},
\label{B8}\\
&&B^8_4=i\qq{D^2\hat DD^2}, \quad
B^8_5=i\qq{D_\mu D_\nu\hat DD_\nu D_\mu},
\nonumber\\
&&B^8_6=i\qq{D_\mu D_\nu\hat DD_\mu D_\nu}, \quad
B^8_7=i\qq{D_\mu\hat D^3D_\mu}.
\nonumber
\end{eqnarray}
Operators similar to those in $B^8_{1\ldots3}$ but with a minus sign
between two terms are $C$--odd, and their vacuum averages vanish.

An interesting new phenomenon similar to the axial anomaly arises for
bilinear quark condensates of dimensions $d\ge8$. At a space dimension
$D\ne4$ we can construct the condensate
\begin{equation}
A=i\qq{D_\al D_\be D_\ga D_\de D_\ep
\ga_{[\al}\ga_\be\ga_\ga\ga_\de\ga_{\ep]}}.
\label{A}
\end{equation}
All coefficients in the expansion of this condensate in gluon condensates
(see Section~\ref{SubCond}) contain traces vanishing in the
4--dimensional space. Momentum integrals in coefficients at $d>8$ gluon
condensates converge, and at $d=8$ they contain divergencies $1/\ep$.
As a result, the quark condensate $A$ is equal to a combination of $d=8$
gluon condensates.

Choosing the anomalous condensate $A$ as one of the basis ones, we can
choose 6 proper quark $d=8$ basis condensates as
\begin{eqnarray}
&&Q^8_1=i\qq{[[G_{\mu\la},G_{\la\nu}]_+,D_\mu]_+\ga_\nu},
\nonumber\\
&&Q^8_2=-\qq{[[G_{\mu\la},\G_{\la\nu}],D_\mu]_+\ga_\nu\ga_5},
\nonumber\\
&&Q^8_3=i\qq{[\hat DG_{\mu\nu},G_{\mu\nu}]}, \quad
Q^8_4=\qq{D^2\hat J},
\label{Q8}\\
&&Q^8_5=i\qq{[G_{\mu\nu},J_\mu]\ga_\nu}, \quad
Q^8_6=\qq{[\G_{\mu\nu},J_\mu]_+\ga_\nu\ga_5}.
\nonumber
\end{eqnarray}
Here again the expressions containing $\G_{\mu\nu}$ and
$\ga_\mu\ga_5=\frac{i}{3!}\ep_{\mu\al\be\ga}\ga_\al\ga_\be\ga_\ga$
are understood as a short notation for the expressions in which both
$\ep$ tensors are eliminated.

All operators in~(\ref{Q8}) (as well as in~(\ref{Q7}), (\ref{Q6})) are
purely $C$--even. $C$--conjugation permutes all $D_\mu$ and $\ga_\mu$
in the opposite order, and changes their signs (because $D_\mu$ is
transformed to $\D_\mu$). Therefore every commutator gives a factor $-1$
to the $C$--parity, and every anticommutator gives $+1$; we should not use
ordinary products if we want to obtain operators with a definite $C$--parity.
According to these rules, $G_{\mu\nu}$ and all its derivatives
$D_\la G_{\mu\nu}$,\dots\ (including $J_\mu$) are $C$--odd; $\ga_\mu$ and
$\si_{\mu\nu}$ are $C$--odd, while 1, $\ga_5$, and $\ga_\mu\ga_5$ are
$C$--even.

The condensates~(\ref{A}), (\ref{Q8}) are expressed via~(\ref{B8}) as
\begin{eqnarray}
&&A=-B^8_1-B^8_2+B^8_3+B^8_4+B^8_5-B^8_6+B^8_7
\nonumber\\
&&\quad{}+2mB^7_1+2mB^7_2-2mB^7_3-2mB^7_4+3m^2B^6-4m^3B^5+m^5B^3,
\nonumber\\
&&Q^8_1=B^8_1-B^8_2+2B^8_5-2B^8_6-2mB^7_1+2mB^7_3,
\nonumber\\
&&Q^8_2=2B^8_1-2B^8_2-2B^8_4+2B^8_5-2B^8_6+2B^8_7
\nonumber\\
&&\quad{}-2mB^7_1+2mB^7_2+2mB^7_3-2mB^7_4-2m^2B^6+2m^3B^5,
\label{QB8}\\
&&Q^8_3=-4B^8_5+4B^8_6+4mB^7_2-4mB^7_3,
\nonumber\\
&&Q^8_4=2B^8_1+2B^8_2-2B^8_4-4B^8_5-2mB^7_1,
\nonumber\\
&&Q^8_5=B^8_1+B^8_2-2B^8_3-2mB^7_1-2mB^7_2+4mB^7_3,
\nonumber\\
&&Q^8_6=3B^8_1+B^8_2-2B^8_3-2B^8_4
\nonumber\\
&&\quad{}-4mB^7_1-2mB^7_2+4mB^7_3-4m^2B^6+4m^3B^5.
\nonumber
\end{eqnarray}
Solving this linear system we obtain
\begin{eqnarray}
&&B^8_1=A+\frac12Q^8_1-\frac12Q^8_2+\frac14Q^8_3-\frac12Q^8_5+Q^8_6
\nonumber\\
&&\quad{}+mQ^7_1-\frac{m}2Q^7_2-2mQ^7_3-m^2Q^6-2m^3Q^5+2m^5Q^3,
\nonumber\\
&&B^8_2=A-\frac12Q^8_1-\frac12Q^8_2-\frac14Q^8_3-\frac12Q^8_5+Q^8_6
\nonumber\\
&&\quad{}+3mQ^7_1-\frac{m}2Q^7_2-2mQ^7_3+mQ^7_4-3m^2Q^6-2m^3Q^5+2m^5Q^3,
\nonumber\\
&&B^8_3=A-\frac12Q^8_2-Q^8_5+Q^8_6
\nonumber\\
&&\quad{}+2mQ^7_1-\frac{m}2Q^7_2-2mQ^7_3+mQ^7_4-3m^2Q^6-2m^3Q^5+2m^5Q^3,
\nonumber\\
&&B^8_4=A+\frac12Q^8_1-\frac12Q^8_2+\frac14Q^8_3+\frac12Q^8_6
+\frac{m}2Q^7_1-mQ^7_3-m^3Q^5+m^5Q^3,
\label{BQ8}\\
&&B^8_5=\frac12A-\frac14Q^8_1-\frac14Q^8_2-\frac18Q^8_3-\frac14Q^8_4
-\frac12Q^8_5+\frac34Q^8_6
\nonumber\\
&&\quad{}+\frac32mQ^7_1-\frac{m}4Q^7_2-mQ^7_3+\frac{m}2Q^7_4-2m^2Q^6
-m^3Q^5+m^5Q^3,
\nonumber\\
&&B^8_6=\frac12A-\frac14Q^8_1-\frac14Q^8_2+\frac18Q^8_3-\frac14Q^8_4
-\frac12Q^8_5+\frac34Q^8_6
\nonumber\\
&&\quad{}+mQ^7_1-\frac{m}4Q^7_2-mQ^7_3+\frac{m}2Q^7_4-2m^2Q^6-m^3Q^5+m^5Q^3,
\nonumber\\
&&B^8_7=A-\frac12Q^8_1+\frac12Q^8_6
\nonumber\\
&&\quad{}+mQ^7_1+\frac{m}2Q^7_2-mQ^7_3-\frac32m^2Q^6-\frac{m^3}2Q^5+m^5Q^3.
\nonumber
\end{eqnarray}

I have implemented the discussed algorithm as a Modula-2 program. It accepts
an arbitrary bilinear quark condensate and expresses it via $B^d_i$
producing a REDUCE readable output, which can be used to express the
condensate via $Q^d_j$ using the formulae~(\ref{BQ6}), (\ref{BQ7}),
(\ref{BQ8}).

\begin{sloppypar}
The problem of four--quark condensates' classification is much more difficult.
Even at $d=6$ there exist infinitely many condensates
$\qq{\ga_{[\mu_1}\ldots\ga_{\mu_n]}}\allowbreak
\qq{\ga_{\mu_1}\ldots\ga_{\mu_n]}}$,
$\qq{t^a\ga_{[\mu_1}\ldots\ga_{\mu_n]}}\allowbreak
\qq{t^a\ga_{\mu_1}\ldots\ga_{\mu_n]}}$.
Those with $n>4$ are anomalous and can be eliminated.
\end{sloppypar}

\subsection{Gluon condensates}
\label{SubGluon}

In this Section we apply similar methods to the classification of gluon
condensates. For each dimension $d=2n$ we introduce a linear space. We shall
call it the extended space. It is generated by the basis of formal sequences
$E^d_i$ constructed from $D_\mu$ in which each index is contained twice.
Sequences obtained from each other by renaming indices, cyclic permutations
(trace cyclicity), and reversing ($C$--parity) are considered equivalent.
Physical gluon condensates can be systematically expressed via this basis,
and form a subspace of the extended space. We can choose a basis in this
subspace composed from the most convenient condensates $G^d_j$, and extend
it to a basis in the whole space using a subset of $E^d_i$. After that
we can expand any condensate in $E^d_i$ and then reexpress it via $G^d_j$
and the selected $E^d_i$. In fact $E^d_i$ should not appear if we have
indeed selected a complete basis in the physical subspace.

General guidelines for choosing good basis condensates are similar
to the quark case. First of all, we should include all independent condensates
containing $J_\mu$. The number of derivatives should be kept minimum.

There is only one gluon condensate with $d=4$, namely
$G^4=\va{\Tr G_{\mu\nu}G_{\mu\nu}}$. The extended space is 2--dimensional:
$E^4_1=D^2D^2$, $E^4_2=D_\mu D_\nu D_\mu D_\nu$; $G^4=2E^4_1-2E^4_2$.
But all this is of no use here.

At $d=6$ the extended space is 5--dimensional:
\begin{eqnarray}
&&E^6_1=D^2D^2D^2, \quad
E^6_2=D^2D_\mu D_\nu D_\mu D_\nu, \quad
E^6_3=D^2D_\mu D^2D_\mu,
\nonumber\\
&&E^6_4=D_\la D_\mu D_\la D_\nu D_\mu D_\nu, \quad
E^6_5=D_\la D_\mu D_\nu D_\la D_\mu D_\nu.
\label{E6}
\end{eqnarray}
There are 2 linearly independent physical condensates:
\begin{equation}
G^6_1=i\va{\Tr G_{\la\mu}G_{\mu\nu}G_{\nu\la}}, \quad
G^6_2=\va{\Tr J_\mu J_\mu}.
\label{G6}
\end{equation}
They are expressed via $E^6_i$ as
\begin{equation}
G^6_1=E^6_1-3E^6_2+3E^6_4-E^6_5, \quad
G^6_2=-2E^6_1+8E^6_2-2E^6_3-4E^6_4.
\label{GE6}
\end{equation}
We can exclude $E^6_{4\ldots5}$:
\begin{eqnarray}
&&E^6_4=-\frac14G^6_2-\frac12E^6_1+2E^6_2-\frac12E^6_3,
\nonumber\\
&&E^6_5=-G^6_1-\frac34G^6_2-\frac12E^6_1+3E^6_2-\frac32E^6_3.
\label{EG6}
\end{eqnarray}
After that we can expand any $d=6$ condensate in~(\ref{E6}) and express it
via~(\ref{G6}) and~$E^6_{1\ldots3}$; the last terms should not appear.
I have verified this for all $d=6$ condensates.

At $d=8$ the extended space is 17--dimensional:
\begin{eqnarray}
&&E^8_1=D^2D^2D^2D^2, \quad
E^8_2=D^2D^2D_\mu D_\nu D_\mu D_\nu,
\nonumber\\
&&E^8_3=D^2D^2D_\mu D^2D_\mu, \quad
E^8_4=D^2D_\la D_\mu D_\la D_\nu D_\mu D_\nu,
\nonumber\\
&&E^8_5=D^2D_\la D_\mu D_\la D^2D_\mu, \quad
E^8_6=D^2D_\la D_\mu D_\nu D_\la D_\mu D_\nu,
\nonumber\\
&&E^8_7=D^2D_\la D_\mu D_\nu D_\la D_\nu D_\mu, \quad
E^8_8=D^2D_\la D_\mu D_\nu D_\mu D_\nu D_\la,
\nonumber\\
&&E^8_9=D^2D_\mu D_\nu D^2D_\mu D_\nu, \quad
E^8_{10}=D^2D_\mu D_\nu D^2D_\nu D_\mu,
\label{E8}\\
&&E^8_{11}=D_\mu D_\nu D_\mu D_\nu D_\ro D_\si D_\ro D_\si, \quad
E^8_{12}=D_\mu D_\nu D_\mu D_\ro D_\nu D_\si D_\ro D_\si,
\nonumber\\
&&E^8_{13}=D_\mu D_\nu D_\mu D_\ro D_\si D_\nu D_\ro D_\si, \quad
E^8_{14}=D_\mu D_\nu D_\mu D_\ro D_\si D_\nu D_\si D_\ro,
\nonumber\\
&&E^8_{15}=D_\mu D_\nu D_\ro D_\mu D_\si D_\nu D_\ro D_\si, \quad
E^8_{16}=D_\mu D_\nu D_\ro D_\mu D_\si D_\ro D_\nu D_\si,
\nonumber\\
&&E^8_{17}=D_\mu D_\nu D_\ro D_\si D_\mu D_\nu D_\ro D_\si.
\nonumber
\end{eqnarray}
A set of independent $d=8$ gluon condensates was found in~\cite{NikRad}.
We choose one condensate in a different way~\cite{GrPin2}:
\begin{eqnarray}
&&G^8_1=\va{\Tr G_{\mu\nu}G_{\mu\nu}G_{\al\be}G_{\al\be}}, \quad
G^8_2=\va{\Tr G_{\mu\nu}G_{\al\be}G_{\mu\nu}G_{\al\be}},
\nonumber\\
&&G^8_3=\va{\Tr G_{\mu\al}G_{\al\nu}G_{\nu\be}G_{\be\mu}}, \quad
G^8_4=\va{\Tr G_{\mu\al}G_{\al\nu}G_{\mu\be}G_{\be\nu}},
\nonumber\\
&&G^8_5=i\va{\Tr J_\mu G_{\mu\nu}J_\nu}, \quad
G^8_6=i\va{\Tr J_\la[D_\la G_{\mu\nu},G_{\mu\nu}]},
\label{G8}\\
&&G^8_7=\va{\Tr J_\mu D^2J_\mu}.
\nonumber
\end{eqnarray}
All condensates are written in an explicitly $C$--even form (even number
of commutators). They are expressed via $E^8_i$ as
\begin{eqnarray}
&&G^8_1=-8E^8_8+4E^8_{10}+4E^8_{11},
\nonumber\\
&&G^8_2=-8E^8_{15}+4E^8_{16}+4E^8_{17},
\nonumber\\
&&G^8_3=E^8_1-4E^8_2+4E^8_4+2E^8_{11}-4E^8_{12}+E^8_{16},
\nonumber\\
&&G^8_4=2E^8_6-4E^8_7+E^8_9+2E^8_{12}-4E^8_{13}+2E^8_{14}+E^8_{15},
\label{GE8}\\
&&G^8_5=E^8_1-5E^8_2+2E^8_3+4E^8_4-2E^8_5+4E^8_7-4E^8_8-E^8_9+E^8_{10}
+4E^8_{11}-4E^8_{12},
\nonumber\\
&&G^8_6=-4E^8_2+4E^8_3+8E^8_6-8E^8_7-12E^8_8+4E^8_{10}+8E^8_{11}-8E^8_{13}
+8E^8_{14},
\nonumber\\
&&G^8_7=-2E^8_1+8E^8_2-6E^8_3-8E^8_4+12E^8_5-16E^8_7+4E^8_9+8E^8_{14}.
\nonumber
\end{eqnarray}
We can exclude $E^8_{11\ldots17}$:
\begin{eqnarray}
&&E^8_{11}=\frac14G^8_1+2E^8_8-E^8_{10},
\nonumber\\
&&E^8_{12}=\frac14G^8_1-\frac14G^8_5
\nonumber\\
&&\quad{}+\frac14E^8_1-\frac54E^8_2+\frac12E^8_3+E^8_4-\frac12E^8_5+E^8_7
+E^8_8-\frac14E^8_9-\frac34E^8_{10},
\nonumber\\
&&E^8_{13}=\frac14G^8_1-\frac18G^8_6+\frac18G^8_7
\nonumber\\
&&\quad{}+\frac14E^8_1-\frac32E^8_2+\frac54E^8_3+E^8_4-\frac32E^8_5+E^8_6
+E^8_7+\frac12E^8_8-\frac12E^8_9-\frac12E^8_{10},
\nonumber\\
&&E^8_{14}=\frac18G^8_7+\frac14E^8_1-E^8_2+\frac34E^8_3+E^8_4-\frac32E^8_5
+2E^8_7-\frac12E^8_9,
\label{EG8}\\
&&E^8_{15}=\frac12G^8_1+G^8_4+\frac12G^8_5-\frac12G^8_6+\frac14G^8_7
\nonumber\\
&&\quad{}-\frac32E^8_2+\frac52E^8_3-2E^8_5+2E^8_6+2E^8_7-\frac32E^8_9
-\frac12E^8_{10},
\nonumber\\
&&E^8_{16}=\frac12G^8_1+G^8_3-G^8_5-E^8_2+2E^8_3-2E^8_5+4E^8_7-E^8_9-E^8_{10},
\nonumber\\
&&E^8_{17}=\frac12G^8_1+\frac14G^8_2-G^8_3+2G^8_4+2G^8_5-G^8_6+\frac12G^8_7
\nonumber\\
&&\quad{}-2E^8_2+3E^8_3-2E^8_5+4E^8_6-2E^8_9.
\nonumber
\end{eqnarray}
After that we can expand any $d=8$ condensate in~(\ref{E8}) and express it
via~(\ref{G8}) and~$E^8_{1\ldots10}$; the last terms should not appear.

I have written a Modula-2 program that expands any gluon condensate
in~$E^d_i$ and produces a REDUCE readable output, which can be used for
expressing this condensate via~$G^d_j$ and selected~$E^d_i$. Using this
program, I have verified that all $d=8$ gluon condensates indeed can
be expanded in the basis~(\ref{G8}).

\subsection{Vacuum averaging}
\label{SubAver}

After obtaining an expression for a correlator via background fields we should
average it over the vacuum. The first possible way is to write down
the most general expressions for vacuum averages with free indices and
to find unknown coefficients by solving linear systems.

For the quark condensates with $d\le6$ we have
\begin{eqnarray}
&&\va{\q_\si q_\ro}=\frac{Q^3}{2^2}1_{\ro\si},
\nonumber\\
&&\va{\q_\si D_\al q_\ro}=-\frac{imQ^3}{2^4}(\ga_\al)_{\ro\si},
\nonumber\\
&&\va{\q_\si D_\al D_\be q_\ro}=\frac1{2^5}
\Bigg[Q^5\left(\de_{\al\be}+\frac{i}3\si_{\al\be}\right)
-2m^2Q^3\de_{\al\be}\Bigg]_{\ro\si},
\label{AvQ}\\
&&\va{\q_\si D_\al D_\be D_\ga q_\ro}=\frac1{2^6 3^3}
\Bigg[Q^6(\de_{\al\be}\ga_\ga+\de_{\be\ga}\ga_\al-5\de_{\al\ga}\ga_\be
-3i\ep_{\al\be\ga\de}\ga_\de\ga_5)
\nonumber\\
&&\quad{}-3mQ^5(\de_{\al\be}\ga_\ga+\de_{\be\ga}\ga_\al+\de_{\al\ga}\ga_\be
-i\ep_{\al\be\ga\de}\ga_\de\ga_5)
\nonumber\\
&&\quad{}+6m^3Q^3(\de_{\al\be}\ga_\ga+\de_{\be\ga}\ga_\al+\de_{\al\ga}\ga_\be)
\Bigg]_{\ro\si}.
\nonumber
\end{eqnarray}
One can easily obtain similar formulae with $G_{\al\be}$ and its derivatives
by antisymmetrization over corresponding indices.

For the gluon condensates with $d\le6$ we have
\begin{eqnarray}
&&\va{g^2G^a_{\mu\nu}G^b_{\ro\si}}=\frac{G^4}{NC_FD(D-1)}\de^{ab}
(\de_{\mu\ro}\de_{\nu\si}-\de_{\mu\si}\de_{\nu\ro}),
\nonumber\\
&&\va{g^2G^a_{\mu\nu}D_\al D_\be G^b_{\ro\si}}=
\frac{\de^{ab}}{NC_FD(D-1)(D^2-4)}
\nonumber\\
&&\quad{}\times\Bigg[-(D+2)G^6_1
(((\de_{\si\mu}\de_{\nu\al}\de_{\be\ro}-(\mu\vv\nu))-(\ro\vv\si))-(\al\vv\be))
\nonumber\\
&&\quad{}+\left[(D-4)G^6_1+(D-2)G^6_2\right]
(((\de_{\si\mu}\de_{\nu\al}\de_{\be\ro}-(\mu\vv\nu))-(\ro\vv\si))+(\al\vv\be))
\nonumber\\
&&\quad{}-\left[4(D-1)G^6_1+2(D-2)G^6_2\right]
\de_{\al\be}(\de_{\mu\ro}\de_{\nu\si}-(\ro\vv\si))\Bigg],
\label{AvG}\\
&&\va{g^3G^a_{\mu\nu}G^b_{\al\be}G^c_{\ro\si}}=-
\frac{2G^6_1}{N^2C_FD(D-1)(D-2)}f^{abc}
\nonumber\\
&&\quad{}\times(((\de_{\si\mu}\de_{\nu\al}\de_{\be\ro}-(\mu\vv\nu))
-(\ro\vv\si))-(\al\vv\be)).
\nonumber
\end{eqnarray}
We can substitute these relations to expressions for correlators and obtain
scalar expressions. But these formulae become very complicated in higher
dimensions.

An alternative and simpler way is to average an expression for a correlator
over $p$ directions. This is done using the recurrent relation ($n$ is even)
\begin{equation}
\overline{p_{\mu_1}p_{\mu_2}\ldots p_{\mu_n}}=\frac{p^2}{D+n-2}
(\de_{\mu_1\mu_2}\overline{p_{\mu_3}\ldots p_{\mu_n}}
+\de_{\mu_1\mu_3}\overline{p_{\mu_2}\ldots p_{\mu_n}}+\cdots
+\de_{\mu_1\mu_n}\overline{p_{\mu_2}\ldots p_{\mu_{n-1}}})
\label{AvR}
\end{equation}
or the explicit formula
\begin{equation}
\overline{p_{\mu_1}p_{\mu_2}\ldots p_{\mu_n}}=
\frac{(p^2)^{n/2}}{D(D+2)\ldots(D+n-2)}
(\de_{\mu_1\mu_2}\de_{\mu_3\mu_4}\ldots\de_{\mu_{n-1}\mu_n}+\cdots),
\label{Av}
\end{equation}
where the sum is taken over all $(n-1)!!$ methods of paring $n$ indices.
Here $D$ is the space dimension; the formulae can be used at any dimension
including non--integer one (the dimensional regularization). After such
an averaging we obtain scalar vacuum condensates for which the methods
of the previous Sections can be directly used.

In calculations of three--point correlators we encounter the averages
$\overline{p_{\mu_1}\ldots p_{\mu_m}q_{\nu_1}\ldots q_{\nu_n}}$
over orientations of the pair $p$, $q$ with a fixed relative orientation.
These averages can be calculated using the decomposition $q=(qp)p/p^2+q_\bot$,
$q_\bot^2=(p^2q^2-(pq)^2)/p^2$. First we average over orientations of $q_\bot$
orthogonal to $p$ in $(D-1)$--dimensional space using~(\ref{Av}) with
$\de_{\mu\nu}\to\de_{\bot\mu\nu}=\de_{\mu\nu}-p_\mu p_\nu/p^2$.
Then we average over orientations of $p$. It is clear that if there is
only one vector $q$ then the averages are given by the formula~(\ref{Av})
in which one $p^2$ is replaced by $pq$. Some less trivial averages are
\begin{eqnarray}
&&\overline{p_\al p_\be q_\mu q_\nu}=\frac1{(D-1)D(D+2)}
\Big[((D+1)p^2q^2-2(pq)^2)\de_{\al\be}\de_{\mu\nu}
\nonumber\\
&&\quad{}+(D(pq)^2-p^2q^2)(\de_{\al\mu}\de_{\be\nu}+\de_{\al\nu}\de_{\be\mu})
\Big],
\nonumber\\
&&\overline{p_\al p_\be p_\ga q_\la q_\mu q_\nu}=\frac{pq}{(D-1)D(D+2)(D+4)}
\nonumber\\
&&\quad{}\times\Big[((D+1)p^2q^2-2(pq)^2)(\de_{\al\be}\de_{\la\mu}\de_{\ga\nu}
+\cdots)
\nonumber\\
&&\quad{}+((D+2)(pq)^2-3p^2q^2)(\de_{\al\la}\de_{\be\mu}\de_{\ga\nu}+\cdots)
\Big],
\label{Avpq}\\
&&\overline{p_\al p_\be p_\ga p_\de q_\mu q_\nu}=\frac{p^2}{(D-1)D(D+2)(D+4)}
\nonumber\\
&&\quad{}\times\Big[((D+3)p^2q^2-4(pq)^2)\de_{\mu\nu}(\de_{\al\be}\de_{\ga\de}
+\cdots)
\nonumber\\
&&\quad{}+(D(pq)^2-p^2q^2)(\de_{\al\be}\de_{\ga\mu}\de_{\de\nu}+\cdots)\Big].
\nonumber
\end{eqnarray}
Of course, these formulae can be also obtained by writing down the most
general forms with unknown coefficients and solving linear systems.

\section{Nonlocal condensates}
\label{SecNonl}\setcounter{equation}{0}

As a sample application we consider expansion of nonlocal vacuum condensates
in local ones. Nonlocal condensates naturally appear in calculations of
correlators~[10--14]; 
their expansion in spacings lead to the usual OPE~(\ref{OPE}). Expansions
derived in this Section can be used either for easy obtaining of OPE of
correlators involving nonlocal condensates, or for inventing consistent
anzatzs for nonlocal condensates.

\subsection{Bilocal condensates}
\label{SubBil}

The gauge--invariant bilocal quark condensate~\cite{MikhRad,Appl}
has the structure
\begin{eqnarray}
&&\va{\q_\si(0)E(0,x)q_\ro(x)}=\frac{\qq{}}{4}
\left[f_S(x^2)-\frac{i\hat x}{D}f_V(x^2)\right]_{\ro\si},
\label{NonQ}\\
&&E(0,x)=P\exp\left(-i\int\limits_0^x A_\mu(y)dy_\mu\right).
\nonumber
\end{eqnarray}
In the fixed--point gauge $E(0,x)=1$, and
\begin{eqnarray}
&&f_S(x^2)=\frac{\va{\q(0)q(x)}}{\qq{}}=
\frac{\qq{\left(1+\frac1{2!}x_\al x_\be D_\al D_\be
+\frac1{4!}x_\al x_\be x_\ga x_\de D_\al D_\be D_\ga D_\de
+\cdots\right)}}{\qq{}}
\nonumber\\
&&{}=1-\frac{B^5}{2!DB^3}x^2
+\frac{B^7_1+B^7_2+B^7_3}{4!D(D+2)B^3}x^4+O(x^6)
\nonumber\\
&&{}=1+\frac{\frac12Q^5-m^2Q^3}{2!DQ^3}x^2
\nonumber\\
&&\quad{}
+\frac{3Q^7_1-\frac32Q^7_2-3Q^7_3+Q^7_4-2mQ^6-3m^2Q^5+3m^4Q^3}{4!D(D+2)Q^3}x^4
+O(x^6),
\label{NonQres}\\
&&f_V(x^2)=\frac{iD\va{\q(0)\hat xq(x)}}{\qq{}x^2}
\nonumber\\
&&{}=\frac{iD\qq{\hat x\left(x_\al D_\al
+\frac1{3!}x_\al x_\be x_\ga D_\al D_\be D_\ga
+\frac1{5!}x_\al x_\be x_\ga x_\de x_\ep D_\al D_\be D_\ga D_\de D_\ep
+\cdots\right)}}{\qq{}x^2}
\nonumber\\
&&{}=m-\frac{B^6+2mB^5}{3!(D+2)B^3}x^2
\nonumber\\
&&\quad{}
+\frac{B^8_1+B^8_2+B^8_3+B^8_4+B^8_5+B^8_6+2m(B^7_1+B^7_2+B^7_3)}
{5!(D+2)(D+4)B^3}x^4+O(x^6)
\nonumber\\
&&{}=m+\frac{\frac12Q^6+\frac32mQ^5-3m^3Q^3}{3!(D+2)Q^3}x^2
\nonumber\\
&&\quad{}+\Big[5A-{\textstyle\frac52}Q^8_2+{\textstyle\frac14}Q^8_3
-{\textstyle\frac12}Q^8_4-3Q^8_5+5Q^8_6+5m(3Q^7_1-Q^7_2-3Q^7_3+Q^7_4)
\nonumber\\
&&\quad{}-15m^2(Q^6+mQ^5-m^3Q^3)\Big]
\frac{x^4}{5!(D+2)(D+4)Q^3}+O(x^6)
\nonumber
\end{eqnarray}
On the first step we use the averaging over $x$ directions~(\ref{Av}) and the
definitions of $B^d_i$~(\ref{B6}), (\ref{B7}), (\ref{B8}), and on the second
step---the expressions for $B^d_i$ via $Q^d_j$~(\ref{BQ6}), (\ref{BQ7}),
(\ref{BQ8}).

The gauge--invariant bilocal gluon condensate~\cite{Simonov,Mikh} has
the structure
\begin{eqnarray}
&&\va{G^a_{\mu\nu}(0) E^{ab}(0,x) G^b_{\rho\si}(x)}
=\frac{\va{G^a_{\mu\nu}G^a_{\mu\nu}}}{D(D-1)} K_{\mu\nu,\rho\si}(x),
\nonumber\\
&&K_{\mu\nu,\rho\si}(x)=D(D-1)\va{\Tr G_{\mu\nu}(0)G_{\ro\si}(x)}/G^4
\nonumber\\
&&{}=(D(x^2)+D_1(x^2))(\de_{\mu\rho}\de_{\nu\si}-(\rho\vv\si))
\label{NonG}\\
&&\quad{}
+\frac{dD_1(x^2)}{dx^2}((x_\mu x_\rho \de_{\nu\si}-(\mu\vv\nu))-(\rho\vv\si)),
\nonumber
\end{eqnarray}
where $E^{ab}(0,x)$ is the $P$--exponent in the adjoint representation equal
to $\de^{ab}$ in the fixed--point gauge. We obtain the linear system
\begin{eqnarray}
&&D(x^2)+D_1(x^2)+\frac{2x^2}{D}D_1'(x^2)=\frac{A(x^2)}{G^4},
\nonumber\\
&&D(x^2)+D_1(x^2)+x^2 D_1'(x^2)=\frac{B(x^2)}{G^4},
\label{LinSys}
\end{eqnarray}
where
\begin{eqnarray}
&&A(x^2)=\va{\Tr G_{\mu\nu}(0)G_{\mu\nu}(x)}
=G^4 + \frac{A_1 x^2}{2D} + \frac{A_2 x^4}{4!D(D+2)}+O(x^6),
\label{AB}\\
&&B(x^2)=\frac{Dx_\mu x_\nu}{x^2} \va{\Tr G_{\mu\la}(0)G_{\nu\la}(x)}
=G^4 + \frac{B_1 x^2}{2(D+2)} + \frac{B_2 x^4}{4!(D+2)(D+4)}+O(x^6).
\nonumber
\end{eqnarray}
Here
\begin{eqnarray}
&&A_1=\va{\Tr G_{\mu\nu} D^2 G_{\mu\nu}},
\nonumber\\
&&B_1=A_1+\va{\Tr G_{\mu\la}(D_\mu D_\nu+D_\nu D_\mu)G_{\nu\la}},
\nonumber\\
&&A_2=\va{\Tr G_{\mu\nu} (D^2 D^2+D_\al D^2 D_\al+D_\al D_\be D_\al D_\be)
G_{\mu\nu}},
\nonumber\\
&&B_2=A_2+\big<\Tr G_{\mu\la}(D_\mu D_\nu D^2+D_\nu D_\mu D^2
+D_\mu D^2 D_\nu+D_\nu D^2 D_\mu
\nonumber\\
&&\quad{}+D^2 D_\mu D_\nu+D^2 D_\nu D_\mu
+D_\mu D_\al D_\nu D_\al+D_\nu D_\al D_\mu D_\al
\nonumber\\
&&\quad{}+D_\al D_\mu D_\nu D_\al+D_\al D_\nu D_\mu D_\al
+D_\al D_\mu D_\al D_\nu+D_\al D_\nu D_\al D_\mu)G_{\nu\la}\big>.
\nonumber
\end{eqnarray}
Using the method described in Section~\ref{SubGluon}, we obtain
\begin{eqnarray}
&&A_1=-4G^6_1-2G^6_2,
\nonumber\\
&&B_1=-6G^6_1-4G^6_2,
\label{ABres}\\
&&A_2=3G^8_{1-2}+24G^8_{3-4}-36G^8_5+10G^8_6-6G^8_7,
\nonumber\\
&&B_2=5G^8_{1-2}+50G^8_{3-4}-84G^8_5+24G^8_6-18G^8_7,
\nonumber
\end{eqnarray}
where $G^8_{1-2}=G^8_1-G^8_2$, $G^8_{3-4}=G^8_3-G^8_4$.
The solution of~(\ref{LinSys}) is
\begin{eqnarray}
&&D(x^2)=\frac1{G^4}\left[\frac{(2A_1-B_1)x^2}{2(D-2)}
+\frac{(3A_2-B_2)x^4}{2\cdot4!(D^2-4)}+O(x^6)\right]
\nonumber\\
&&\quad{}=-\frac{G^6_1}{(D-2)G^4}x^2
+\frac{2G^8_{1-2}+11G^8_{3-4}-12G^8_5+3G^8_6}{(D^2-4)G^4}\frac{x^4}{4!}
+O(x^6),
\label{NonGres}\\
&&D_1(x^2)=1+\frac1{G^4}\frac{D}{D-2}\left[
\left(\frac{B_1}{D+2}-\frac{A_1}{D}\right)\frac{x^2}{2}
+\left(\frac{B_2}{D+4}-\frac{A_2}{D}\right)\frac{x^4}{2\cdot4!(D+2)}
+O(x^6)\right]
\nonumber\\
&&\quad{}=1-\frac{(D-4)G^6_1+(D-2)G^6_2}{(D^2-4)G^4}x^2
\nonumber\\
&&\quad{}+\Big[(D-6)G^8_{1-2}+(13D-48)G^8_{3-4}-24(D-3)G^8_5+(7D-20)G^8_6
\nonumber\\
&&\quad{}-6(D-2)G^8_7\Big]\frac{x^4}{4!(D+4)(D^2-4)G^4}+O(x^6).
\nonumber
\end{eqnarray}
In the abelian case, only $D_1(x^2)$ survives~\cite{Simonov};
this is the reason why $D(x^2)$ (\ref{NonGres}) contains only
purely nonabelian condensates $G^6_1$, $G^8_{1-2}$, $G^8_{3-4}$, $G^8_{5,6}$.

\subsection{Noncollinear condensates}
\label{SubNonc}

Three--point correlators, sum rules for hadron wave functions, and some other
kinds of calculations naturally involve trilocal objects which we shall
call noncollinear condensates. The noncollinear quark condensate has
the structure
\begin{eqnarray}
&&\va{\q_\si(y)E(y,0)E(0,x)q_\ro(x)}
\label{NcQ}\\
&&\quad{} = \frac{\qq{}}4 \left[ f_S(x,y)
+ \frac{Q^5}{4D(D-1)Q^3}f_T(x,y)[\hat x,\hat y]
- \frac{i}D \left(f_V(x,y)\hat{x}-f_V(y,x)\hat{y}\right) \right]_{\ro\si}.
\nonumber
\end{eqnarray}
In the fixed--point gauge $E(y,0)=E(0,x)=1$. The difference
\begin{eqnarray}
&&f_S(x,y)-f_S((x-y)^2) = \frac{\va{\q(y)q(x)}-\va{\q(0)q(x-y)}}{\qq{}}
\nonumber\\
&&\quad{} = \frac{\Delta\qq{(2D^2D^2-D_\mu D^2D_\mu-D_\mu D_\nu D_\mu D_\nu)}}
{12D(D-1)\qq{}} + O(x^3,y^3)
\label{NcQS}\\
&&\quad = - \frac{\Delta\left(3Q^7_1+2Q^7_4-4mQ^6\right)}{24D(D-1)Q^3}
+ O(x^3,y^3)
\nonumber
\end{eqnarray}
(where $\Delta=x^2y^2-(xy)^2$) vanishes when $x$ and $y$ are collinear,
as expected. The coefficient $f_T(x,y)$ multiplies the structure
$[\hat x,\hat y]$ vanishing in the collinear limit, therefore it need not
vanish itself:
\begin{eqnarray}
&&f_T(x,y) = \frac{D(D-1)\va{\q(y)[\hat y,\hat x]q(x)}}{Q^5\Delta} = 1
\label{NcQT}\\
&&\quad{} + \frac{\left(6Q^7_1-3Q^7_2-6Q^7_3+2Q^7_4
-3mQ^6-3m^2Q^5\right)(x-y)^2 - \left(3Q^7_3+2Q^7_4\right)xy}
{6(D+2)Q^5} + \cdots
\nonumber
\end{eqnarray}
For the vector matrix elements we find
\begin{eqnarray}
&&i\va{\q(y)\hat xq(x)}=-i\va{\q(y)\hat yq(x)}_{x\vv y}
\nonumber\\
&&\quad{} = \frac{Q^3(x^2-xy)}{D}f_V((x-y)^2)
- \frac{\Delta}{12D(D-1)}g(x,y)
\label{NcQV0}
\end{eqnarray}
where
\begin{eqnarray}
&&g(x,y) = Q^6 + \frac1{40(D+2)} \Big[
-20A(x^2-y^2)-30Q^8_1(x^2-xy)+10Q^8_2(x^2-y^2)
\nonumber\\
&&\quad{}-Q^8_3(13x^2-y^2-9xy)-2Q^8_4(2x^2+y^2-6xy)-2Q^8_5(7x^2+6y^2-11xy)
\nonumber\\
&&\quad{}+20Q^8_6(y^2-xy)+60mQ^7_1(x^2-xy)-10mQ^7_2(x^2-y^2)+40mQ^7_4(x^2-xy)
\nonumber\\
&&\quad{}-20m^2Q^6(5x^2+y^2-6xy)\Big] + O(x^3,y^3).
\label{NcQV1}
\end{eqnarray}
Solving the system of two linear equations, we finally obtain
\begin{equation}
f_V(x,y) = f_V((x-y)^2) - \frac{y^2g(x,y)+xy\, g(y,x)}{12(D-1)Q^3}.
\label{NcQV}
\end{equation}

The noncollinear gluon condensate has the structure
\begin{eqnarray}
&&\va{G^a_{\mu\nu}(y) E^{ab}(y,0) E^{bc}(0,x) G^c_{\rho\si}(x)}
=\frac{\va{G^a_{\mu\nu}G^a_{\mu\nu}}}{D(D-1)} K_{\mu\nu,\rho\si}(x,y),
\nonumber\\
&&T_{\mu\nu\ro\si}
= G^4\left(K_{\mu\nu,\ro\si}(x,y)-K_{\mu\nu,\ro\si}(x-y)\right)
\label{NcG}\\
&&\quad{} = c_1(\de_{\mu\ro}\de_{\nu\si}-(\ro\vv\si))
\nonumber\\
&&\quad{} + c_2((x_\mu x_\ro\de_{\nu\si}-(\mu\vv\nu))-(\ro\vv\si))
\nonumber\\
&&\quad{} + c_3((y_\mu y_\ro\de_{\nu\si}-(\mu\vv\nu))-(\ro\vv\si))
\nonumber\\
&&\quad{} + c_4((x_\mu y_\ro\de_{\nu\si}-(\mu\vv\nu))-(\ro\vv\si))
\nonumber\\
&&\quad{} + c_5((y_\mu x_\ro\de_{\nu\si}-(\mu\vv\nu))-(\ro\vv\si))
\nonumber\\
&&\quad{} + c_6(x_\mu y_\nu-(\mu\vv\nu))(x_\ro y_\si-(\ro\vv\si))
\nonumber
\end{eqnarray}
with $c_3(x,y)=c_2(y,x)$. We find
\begin{eqnarray}
&&T_{\mu\nu,\mu\nu} = \frac{\Delta}{12}\left(-3G^8_{1-2}+2G^8_6\right),
\nonumber\\
&&T_{\la\mu,\la\nu} x_\mu x_\nu
= \left(T_{\la\mu,\la\nu} y_\mu y_\nu\right)_{x\vv y}
\label{NcG1}\\
&&\quad{} = \frac{\Delta}{12(D+2)} \left[(4x^2+xy)\left(-G^8_{1-2}+G^8_6\right)
-(x^2-2xy)G^8_{3-4}-4xy\,G^8_5\right],
\nonumber\\
&&T_{\la\mu,\la\nu} x_\mu y_\nu = \frac{\Delta}{12(D+2)} \Big[12(D+2)G^6_1
-(2x^2+2y^2+xy)G^8_{1-2}
\nonumber\\
&&\quad{}-(11x^2+11y^2-20xy)G^8_{3-4}
+12(x-y)^2G^8_5-(3x^2+3y^2-10xy)G^8_6\Big],
\nonumber\\
&&T_{\la\mu,\la\nu} y_\mu x_\nu = \frac{\Delta}{12(D+2)} \Big[-12(D+2)G^6_1
+(x^2+y^2-7xy)G^8_{1-2}
\nonumber\\
&&\quad{}+(13x^2+13y^2-22xy)G^8_{3-4}
-8(2x^2+2y^2-3xy)G^8_5+2(2x^2+2y^2-xy)G^8_6\Big],
\nonumber\\
&&T_{\mu\nu,\ro\si} x_\mu x_\ro y_\nu y_\si
= \frac{\Delta^2}{12(D+1)(D+2)} \left(-5G^8_{1-2}+4G^8_{3-4}+12G^8_6\right).
\nonumber
\end{eqnarray}
Solving the system of 6 linear equations, we finally obtain
\begin{eqnarray}
&&c_1 = \frac{\Delta\left[-D(3D-7)G^8_{1-2}+4(D+3)G^8_{3-4}
+2(D-2)(D-3)G^8_6\right]}{12(D+1)(D+2)(D-2)(D-3)},
\nonumber\\
&&c_2 = \frac1{12(D+1)(D+2)(D-2)(D-3)} \Big\{ 4(D+1)(D-3)xyG^8_5
\nonumber\\
&&\quad{} + (D-3)\left[D(2y^2-xy)-4y^2-xy\right]G^8_6
\nonumber\\
&&\quad{} - \left[D^2(y^2-xy)-2D(3y^2-xy)+3(y^2+xy)\right]G^8_{1-2}
\nonumber\\
&&\quad{} - \left[(D^2-3)(y^2+2xy)+2D(3y^2-2xy)\right]G^8_{3-4} \Big\},
\nonumber\\
&&c_3 = c_4(x\vv y),
\label{NcGc}\\
&&c_4-c_5 = \frac1{12(D^2-4)} \Big[24(D+2)G^6_1-3(x-y)^2G^8_{1-2}
\nonumber\\
&&\quad{}-6(4x^2+4y^2-7xy)G^8_{3-4}+(7x^2+7y^2-12xy)(4G^8_5-G^8_6)\Big],
\nonumber\\
&&c_4+c_5 = \frac1{12(D+1)(D+2)(D-2)(D-3)} \Big\{ -4(D+1)(D-3)(x^2+y^2)G^8_5
\nonumber\\
&&\quad{} + (D-3)\left[D(x^2+y^2-4xy)+x^2+y^2+8xy\right]G^8_6
\nonumber\\
&&\quad{} - \left[D^2(x-y)^2-2D(x^2+y^2-6xy)-3(x+y)^2\right]G^8_{1-2}
\nonumber\\
&&\quad{} + 2\left[(D^2-3)(x^2+y^2+xy)-2D(x^2+y^2-3xy)\right]G^8_{3-4}
\Big\},
\nonumber\\
&&c_6 = \frac{(D-4)G^8_{1-2}+(D-1)^2G^8_{3-4}+(D-2)(D-3)G^8_6}
{2(D+1)(D+2)(D-2)(D-3)}.
\nonumber
\end{eqnarray}
In the abelian case $K_{\mu\nu,\ro\si}(x,y)=K_{\mu\nu,\ro\si}(x-y)$
has only the $D_1((x-y)^2)$ term~(\ref{NonGres}), therefore
$T_{\mu\nu,\ro\si}$ contains only purely nonabelian condensates.

All calculations in this Section were performed using REDUCE
(with some REDUCE code generated by the Modula-2 programs described
in Section~\ref{SecCond}). Some of them require a lot of memory:
I could not complete the gluonic calculations on a PC with 8 megabytes.

\subsection{Expansion in bilocal condensates}
\label{SubColl}

In some applications, such as sum rules for wave function moments,
it is convenient to expand noncollinear condensates in series in $x$
keeping $y$ finite. Coefficients in these series are bilocal condensates
(Section~\ref{SubBil}).

Expanding a noncollinear quark condensate $\va{\q(y)E(y,0)\Gamma E(0,x)q(x)}$
in $x$, we obtain terms like $\va{\q(y)E(y,0)Oq(0)}$ where $O$ involves
derivatives and gluon fields at $x=0$. Therefore we encounter the problem
of classification of such bilocal condensates. Similarly to Section~%
\ref{SubQuark}, we can easily reduce any condensate of this kind to
a linear combination of terms with $O$ constructed from $\hat D$, $(yD)$,
and $D_\mu$, maybe multiplied by $\hat y$ from the left. Terms with
$\hat D$ at the right reduce to lower--dimensional condensates multiplied
by $m$. Terms with $(yD)$ at the left reduce to derivatives of lower--%
dimensional condensates: expanding the equality
$\va{\q(y)E(y,0)(Oq)_{x=\alpha y}}
=\va{\q((1-\alpha)y)E((1-\alpha)y,0)(Oq)_{x=0}}$ in $\alpha$ we obtain
\begin{equation}
\va{\q(y)E(y,0)(yD)Oq(0)} = -2y^2\frac{d}{dy^2} \va{\q(y)E(y,0)Oq(0)}.
\label{Deriv}
\end{equation}

There are 2 basis dimension 3 bilocal condensates:
\begin{equation}
B^3_1(y^2)=\va{\q(y)E(y,0)q(0)}, \quad
B^3_2(y^2)=-i\va{\q(y)E(y,0)\hat{y}q(0)}.
\label{BiB3}
\end{equation}
They can be simply expressed via 2 bilocal functions considered in
Section~\ref{SubBil}:
\begin{equation}
B^3_1(y^2)=Q^3f_S(y^2), \quad
B^3_2(y^2)=Q^3f_V(y^2)\frac{y^2}{D}.
\label{BiQ3}
\end{equation}
Expansions of these functions at small $y^2$ are given by equation~%
(\ref{NonQres}).

There are 4 basis dimension 5 bilocal condensates:
\begin{eqnarray}
&&B^5_1(y^2) = -\va{\q(y)E(y,0)D^2q(0)},
\nonumber\\
&&B^5_2(y^2) = \va{\q(y)E(y,0)\hat{y}\hat{D}(yD)q(0)},
\nonumber\\
&&B^5_3(y^2) = i\va{\q(y)E(y,0)\hat{y}D^2q(0)},
\label{BiB5}\\
&&B^5_4(y^2) = i\va{\q(y)E(y,0)\hat{D}(yD)q(0)}.
\nonumber
\end{eqnarray}
Expanding in $y$ and averaging over $y$ directions (Section~\ref{SubAver})
we obtain the series
\begin{eqnarray}
&&B^5_1(y^2) = B^5 - \frac{B^7_1 y^2}{2D} + O(y^4),
\nonumber\\
&&B^5_2(y^2) = \frac{-2B^5+m^2B^3}{D}y^2
+ \frac{2B^7_1+B^7_4+mB^6-m^2B^5}{2D(D+2)}y^4 + O(y^6),
\nonumber\\
&&B^5_3(y^2) = \frac{mB^5}{D}y^2 - \frac{B^8_1+2B^8_4+2mB^7_1}{12D(D+2)}y^4
+ O(y^6),
\label{BiB5e}\\
&&B^5_4(y^2) = \frac{B^6}{D}y^2 - \frac{B^8_1+B^8_2+B^8_3}{12D(D+2)}y^4
+ O(y^6).
\nonumber
\end{eqnarray}
We introduce 4 new bilocal functions
\begin{eqnarray}
&&Q^5 f_1(y^2) = i\va{\q(y)E(y,0)G_{\mu\nu}(0)\si_{\mu\nu}q(0)},
\nonumber\\
&&Q^5 f_2(y^2) = \frac{iDy_\mu y_\nu}{y^2}
\va{\q(y)E(y,0)G_{\la\mu}(0)\si_{\la\nu}q(0)},
\nonumber\\
&&Q^6 f_3(y^2) = \frac{2D}{y^2} \va{\q(y)E(y,0)y_\mu G_{\mu\nu}(0)\ga_\nu
q(0)},
\label{BiQ5}\\
&&Q^6 f_4(y^2) = \frac{2iD}{y^2} \va{\q(y)E(y,0)y_\mu\G_{\mu\nu}(0)\ga_\nu
q(0)}.
\nonumber
\end{eqnarray}
They are expressed via the basis bilocal condensates~(\ref{BiB5}) as
\begin{eqnarray}
&&Q^5 f_1(y^2) = -2B^5_1(y^2) + 2m^2B^3f_S(y^2),
\nonumber\\
&&Q^5 f_2(y^2) = \frac{D}{y^2}B^5_2(y^2)
+ mB^3\left(f_V(y^2)+2y^2f'_V(y^2)\right),
\nonumber\\
&&Q^6 f_3(y^2) = -\frac{2D}{y^2}B^5_4(y^2) - 4DmB^3f'_S(y^2),
\label{BiQB5}\\
&&Q^6 f_4(y^2) = \frac{2D}{y^2}\left(B^5_3(y^2)-B^5_4(y^2)\right)
- 4DmB^3f'_S(y^2) - 2m^2B^3f_V(y^2).
\nonumber
\end{eqnarray}
Solving this linear system we obtain
\begin{eqnarray}
&&B^5_1(y^2) = -\frac12Q^5f_1(y^2) + m^2Q^3f_S(y^2),
\nonumber\\
&&B^5_2(y^2) = \frac{y^2}{D} \left[Q^5f_2(y^2)
- mQ^3\left(f_V(y^2)+2y^2f'_V(y^2)\right)\right],
\nonumber\\
&&B^5_3(y^2) = \frac{y^2}{2D} \left[Q^6\left(f_4(y^2)-f_3(y^2)\right)
+2m^2Q^3f_V(y^2)\right],
\label{BiBQ5}\\
&&B^5_4(y^2) = -\frac{y^2}{2D} \left[Q^6f_3(y^2) + 4DmQ^3f'_S(y^2)\right].
\nonumber
\end{eqnarray}
Substituting~(\ref{BiB5e}) and~(\ref{NonQres}) into~(\ref{BiQB5})
we arrive at the small $y^2$ expansions
\begin{eqnarray}
&&f_1(y^2) = 1 + \frac{Q^7_1-Q^7_2-2Q^7_3-m^2Q^5}{2DQ^5}y^2 + O(y^4),
\nonumber\\
&&f_2(y^2) = 1 + \frac{2Q^7_1-Q^7_2-3Q^7_3-mQ^6-m^2Q^5}{2(D+2)Q^5}y^2
+ O(y^4),
\label{BiQ5e}\\
&&f_3(y^2) = 1 + \frac{6A-3Q^8_2-4Q^8_5+6Q^8_6+3mQ^7_2-6m^2Q^6}{12(D+2)Q^6}y^2
+ O(y^4),
\nonumber\\
&&f_4(y^2) = 1 - \frac{mQ^5}{Q^6} + \Big[-6Q^8_1-3Q^8_3-6Q^8_5+4Q^8_6
\nonumber\\
&&\quad{}+12m\left(-Q^7_1+Q^7_2+2Q^7_3-mQ^6+m^2Q^3\right)\Big]
\frac{y^2}{24(D+2)Q^6} + O(y^4).
\nonumber
\end{eqnarray}

Now it is easy to expand the noncollinear quark condensate~(\ref{NcQ})
in small $x$ keeping $y$ fixed. We expand $q(x)$, substitute
$x=(xy/y^2)y+x_\bot$ (with $x_\bot^2=\Delta/y^2$), and perform the
$(D-1)$--dimensional averaging (Section~\ref{SubAver}) over $x_\bot$
directions orthogonal to $y$. Then we express the result in terms of
the basis bilocal condensates~(\ref{BiQ3}), (\ref{BiQ5}), and substitute~%
(\ref{BiQ3}), (\ref{BiBQ5}). Finally, we arrive at the result
\begin{eqnarray}
&&f_S(x,y) = f_S(y^2) - 2(xy)f'_S(y^2)
+ \frac{Q^5\Delta}{4(D-1)Q^3y^2}f_1(y^2)
\nonumber\\
&&\quad{}
- \frac{x^2y^2-D(xy)^2}{(D-1)y^2}\left(f'_S(y^2)+2y^2f''_S(y^2)\right)
- \frac{m^2\Delta}{2(D-1)y^2}f_S(y^2) + O(x^3)
\nonumber\\
&&\quad = f_S((y-x)^2) + \frac{\Delta}{(D-1)y^2}
\left[\frac14\frac{Q^5}{Q^3}f_1(y^2) -Df'_S(y^2) - 2y^2f''_S(y^2)
- \frac{m^2}{2}f_S(y^2)\right] + O(x^3),
\nonumber\\
&&f_T(x,y) = 4D\frac{Q^3}{Q^5} \Bigg\{f'_S(y^2) + \frac{m}{2D}f_V(y^2)
\label{NcQx}\\
&&\quad{} - \frac{xy}{y^2} \left[-\frac{1}{4D}\frac{Q^5}{Q^3}f_2(y^2)
+ f'_S(y^2) + 2y^2f''_S(y^2) + \frac{m}{2D}\left(f_V(y^2)+2y^2f'_V(y^2)\right)
\right]\Bigg\} + O(x^2),
\nonumber\\
&&g(y,x) = 3Q^6\left(f_4(y^2)-f_3(y^2)\right)
+ 6Q^3\left[2(D+2)f'_V(y^2)+4y^2f''_V(y^2)+m^2f_V(y^2)\right] + O(x),
\nonumber\\
&&g(x,y) = \frac{12Q^3}{y^2} \left(Df_V(y^2)+2y^2f'_V(y^2)-Dmf_S(y^2)\right)
\nonumber\\
&&\quad{} - \frac{3xy}{y^2} \Big\{Q^6\left(f_4(y^2)-2f_3(y^2)\right)
\nonumber\\
&&\quad{}
+ 2Q^3\left[6(D+2)f'_V(y^2)+12y^2f''_V(y^2)-4Dmf'_S(y^2)+m^2f_V(y^2)\right]
\Big\} + O(x^2).
\nonumber
\end{eqnarray}
If $y$ is also small, we substitute the $y^2$ expansions of the bilocal
condensates~(\ref{NonQres}), (\ref{BiQ5e}) and obtain expansions consistent
with~(\ref{NcQS}), (\ref{NcQT}), (\ref{NcQV1}). This provides a strong
check of our results.

Expanding a noncollinear gluon condensate in $x$,
we obtain terms like $\va{\Tr G_{\mu\nu}(y,0)O_{\mu\nu}(0)}$
where $G_{\mu\nu}(y,0)=gG^a_{\mu\nu}E^{ab}t^b$
and $O_{\mu\nu}(0)$ is constructed from $G_{\mu\mu}(0)$ and $D_\mu$.
A straightforward generalization of the systematic classification procedure
of Section~\ref{SubGluon} to the case of such bilocal condensates seems
impossible. Fortunately, it is not difficult to find all basis bilocal
condensates up to dimension~6. Two dimension~4 condensates were
introduced in Section~\ref{SubBil}:
\begin{equation}
A(y^2) = \va{\Tr G_{\mu\nu}(y,0)G_{\mu\nu}(0)}, \quad
B(y^2) = \va{\Tr G_{\la\mu}(y,0)G_{\la\nu}(0)} \frac{Dy_\mu y_\nu}{y^2}.
\label{BiG4}
\end{equation}
There is one dimension~5 bilocal condensate
\begin{equation}
C(y^2) = \va{\Tr G_{\mu\nu}(y,0)J_\mu(0)} \frac{Dy_\nu}{y^2},
\label{BiG5}
\end{equation}
and~5 dimension~6 ones
\begin{eqnarray}
&&F_1(y^2) = i\va{\Tr G_{\mu\nu}(y,0)G_{\nu\la}(0)G_{\la\mu}(0)},
\nonumber\\
&&F_2(y^2) = i\va{\Tr G_{\mu\nu}(y,0)G_{\nu\ro}(0)G_{\si\mu}(0)}
\frac{Dy_\ro y_\si}{y^2},
\nonumber\\
&&F_3(y^2) = i\va{\Tr G_{\mu\ro}(y,0)G_{\si\la}(0)G_{\la\mu}(0)}
\frac{Dy_\ro y_\si}{y^2}
\nonumber\\
&&\quad{} = i\va{\Tr G_{\ro\nu}(y,0)G_{\nu\la}(0)G_{\la\si}(0)}
\frac{Dy_\ro y_\si}{y^2},
\label{BiG6}\\
&&F_4(y^2) = \va{\Tr G_{\mu\nu}(y,0)D_\mu J_\nu(0)},
\nonumber\\
&&F_5(y^2) = \va{\Tr G_{\mu\ro}(y,0)D_\mu J_\si(0)} \frac{Dy_\ro y_\si}{y^2}
\nonumber
\end{eqnarray}
(the equivalence of the two definitions of $F_3(y^2)$ follows from the
$C$--conjugation symmetry). Small $y^2$ expansions~(\ref{AB}), (\ref{ABres})
of $A(y^2)$, $B(y^2)$ were obtained in Section~\ref{SubBil}; the other
expansions are
\begin{eqnarray}
&&C(y^2) = G^6_2 + \frac{8G^8_5-3G^8_6+6G^8_7}{12(D+2)}y^2 + O(y^4),
\nonumber\\
&&F_1(y^2) = G^6_1 + \frac{-4G^8_{3-4}+4G^8_5-G^8_6}{4D}y^2 + O(y^4),
\nonumber\\
&&F_2(y^2) = G^6_1 + \frac{-G^8_{1-2}-4G^8_{3-4}+4G^8_5-G^8_6}{4(D+2)}y^2
+ O(y^4),
\nonumber\\
&&F_3(y^2) = G^6_1 + \frac{-3G^8_{3-4}+4G^8_5-G^8_6}{2(D+2)}y^2 + O(y^4),
\label{BiGe}\\
&&F_4(y^2) = G^6_2 + \frac{2G^8_5-G^8_6+G^8_7}{2D}y^2 + O(y^4),
\nonumber\\
&&F_5(y^2) = G^6_2 + \frac{-3G^8_6+2G^8_7}{4(D+2)}y^2 + O(y^4).
\nonumber
\end{eqnarray}

Now we can derive the expansions in the bilocal gluon condensates
\begin{eqnarray}
&&T_{\mu\nu,\mu\nu} = -\frac{D\Delta}{y^2}
\left(2F_1(y^2)+F_4(y^2)+DA'(y^2)+2y^2A''(y^2)\right) + O(x^3),
\nonumber\\
&&T_{\la\mu,\la\nu}y_\mu y_\nu =
-\Delta \Bigg[2F_3(y^2)+\frac12F_5(y^2)+\frac12C(y^2)+y^2C'(y^2)
\nonumber\\
&&\quad{}+DB'(y^2)+2y^2B''(y^2)+D\frac{A(y^2)-B(y^2)}{y^2}\Bigg] + O(x^3),
\nonumber\\
&&T_{\la\mu,\la\nu}y_\mu x_\nu =
\Delta \left(C(y^2)+2B'(y^2)-D\frac{A(y^2)-B(y^2)}{y^2}\right)
\nonumber\\
&&\quad{} + \frac{\Delta xy}{y^2}
\Bigg[-3F_3(y^2)-\frac12F_5(y^2)-\frac32C(y^2)-3y^2C'(y^2)
\nonumber\\
&&\qquad{}-2D\frac{A(y^2)-B(y^2)}{y^2}+2DA'(y^2)
-(3D+2)B'(y^2)-6y^2B''(y^2)\Bigg] + O(x^4),
\nonumber\\
&&T_{\la\mu,\la\nu}x_\mu y_\nu =
\Delta \left(2B'(y^2)-DA'(y^2)-D\frac{A(y^2)-B(y^2)}{y^2}\right)
\label{NcGx}\\
&&\quad{} + \frac{\Delta xy}{y^2}
\Bigg[F_2(y^2)-2F_3(y^2)-\frac12F_5(y^2)-\frac12C(y^2)-y^2C'(y^2)
-2D\frac{A(y^2)-B(y^2)}{y^2}
\nonumber\\
&&\qquad{}+3DA'(y^2)-(3D+2)B'(y^2)+2Dy^2A''(y^2)-6y^2B''(y^2)\Bigg]
+ O(x^4),
\nonumber\\
&&T_{\la\mu,\la\nu}x_\mu x_\nu = \frac{\Delta xy}{y^2}
\left(C(y^2)-DA'(y^2)+4B'(y^2)-2D\frac{A(y^2)-B(y^2)}{y^2}\right)
\nonumber\\
&&\quad{} + \frac{\Delta}{(D+1)y^4} \Bigg[
- D\Delta\left(3F_1(y^2)+2F_4(y^2)\right)
\nonumber\\
&&\qquad{}+\frac12\left(x^2y^2-(D+2)(xy)^2\right)
\left(-2F_2(y^2)+6F_3(y^2)+F_5(y^2)+3C(y^2)+6y^2C'(y^2)\right)
\nonumber\\
&&\qquad{} + D(D+1)\left(x^2y^2-4(xy)^2\right)\frac{A(y^2)-B(y^2)}{y^2}
\nonumber\\
&&\qquad{} - D\left((D+3)x^2y^2-2(3D+4)(xy)^2\right)A'(y^2)
\nonumber\\
&&\qquad{} + \left((D+4)x^2y^2-(5D^2+10D+8)(xy)^2\right)B'(y^2)
\nonumber\\
&&\qquad{} - 2D\left(2x^2y^2-(D+3)(xy)^2\right)y^2A''(y^2)
\nonumber\\
&&\qquad{} + 2\left(3x^2y^2-(5D+8)(xy)^2\right)y^2B''(y^2)
\Bigg] + O(x^5),
\nonumber\\
&&T_{\mu\nu,\ro\si}y_\mu x_\nu y_\ro x_\si =
- \frac{\Delta^2}{2(D+1)y^2} \Big[6F_3(y^2)+3F_5(y^2)+C(y^2)+2y^2C'(y^2)
\nonumber\\
&&\quad{}+2(D+2)B'(y^2)+4y^2B''(y^2)\Big] + O(x^5).
\nonumber
\end{eqnarray}
Solving the system of 6 linear equations, we finally obtain
\begin{eqnarray}
&&c_1 = \frac{\Delta}{(D+1)(D-2)y^2} \Bigg[
\frac{2(D-2)(-DF_1(y^2)+2F_3(y^2))+2F_2(y^2)}{D-3}
\nonumber\\
&&\quad{}-DF_4(y^2)+F_5(y^2)+C(y^2)+2y^2C'(y^2)
\nonumber\\
&&\quad{}-(D+2)(DA'(y^2)-2B'(y^2))
-2y^2(DA''(y^2)-2B''(y^2)) + O(x) \Bigg],
\nonumber\\
&&c_2 = \frac{1}{(D+1)(D-2)} \Bigg[
\frac{(D-5)(-DF_1(y^2)+2F_3(y^2))-(D-1)F_2(y^2)}{D-3}
\nonumber\\
&&\quad{}-DF_4(y^2)+F_5(y^2)+C(y^2)+2y^2C'(y^2)+D(D+1)\frac{A(y^2)-B(y^2)}{y^2}
\nonumber\\
&&\qquad{}-DA'(y^2)+2B'(y^2)-2y^2(DA''(y^2)-2B''(y^2)) + O(x) \Bigg],
\nonumber\\
&&c_3 = \frac{1}{D-2} \Bigg\{ \frac{xy}{y^2}
\left(-C(y^2)+2D\frac{A(y^2)-B(y^2)}{y^2}+DA'(y^2)-4B'(y^2)\right)
\nonumber\\
&&\quad{} + \frac{1}{2(D+1)} \frac{x^2}{y^2} \Bigg[
\frac{4D(D-1)F_1(y^2)-4F_2(y^2)-2(D+1)(2D-5)F_3(y^2)}{D-3}
\nonumber\\
&&\qquad{}+2DF_4(y^2)-DF_5(y^2)-(D+2)(C(y^2)+2y^2C'(y^2))
\nonumber\\
&&\qquad{}-2D(D+1)\frac{A(y^2)-B(y^2)}{y^2}
+2D(D+2)A'(y^2)-2(D^2+2D+2)B'(y^2)
\nonumber\\
&&\qquad{}+4Dy^2A''(y^2)-4(D+2)y^2B''(y^2)
\Bigg]
\label{NcGcx}\\
&&\quad{} + \frac{1}{2(D+1)} \frac{(xy)^2}{y^4} \Bigg[
-2D(3F_1(y^2)+2F_4(y^2))+(D+2)(-2F_2(y^2)
\nonumber\\
&&\qquad{}+6F_3(y^2)+F_5(y^2)+3C(y^2)+6y^2C'(y^2))
+8D(D+1)\frac{A(y^2)-B(y^2)}{y^2}
\nonumber\\
&&\qquad{}-4D(3D+4)A'(y^2)+2(5D^2+10D+8)B'(y^2)
\nonumber\\
&&\qquad{}-4D(D+3)y^2A''(y^2)+4(5D+8)y^2B''(y^2)
\Bigg] + O(x^3) \Bigg\},
\nonumber\\
&&c_4-c_5 = \frac{1}{D-2} \Bigg[-C(y^2)-DA'(y^2)
+ \frac{xy}{y^2} \Big(F_2(y^2)+F_3(y^2)
\nonumber\\
&&\quad{}+C(y^2)+2y^2C'(y^2)+D(A'(y^2)+2y^2A''(y^2))\Big)
+ O(x^2) \Bigg],
\nonumber\\
&&c_4+c_5 = \frac{1}{D-2} \Bigg\{
C(y^2)-2D\frac{A(y^2)-B(y^2)}{y^2}-DA'(y^2)+4B'(y^2)
\nonumber\\
&&\quad{} + \frac{1}{D+1} \frac{xy}{y^2} \Bigg[
\frac{2D(D-5)F_1(y^2)+(D^2-5)F_2(y^2)-(D^2+2D-23)F_3(y^2)}{D-3}
\nonumber\\
&&\qquad{}+2DF_4(y^2)-2F_5(y^2)-(D+3)(C(y^2)+2y^2C'(y^2))
\nonumber\\
&&\qquad{}-4D(D+1)\frac{A(y^2)-B(y^2)}{y^2}
+D(5D+7)A'(y^2)-4(D^2+2D+2)B'(y^2)
\nonumber\\
&&\qquad{}+2D(D+3)y^2A''(y^2)-8(D+2)y^2B''(y^2)
\Bigg] + O(x^2) \Bigg\},
\nonumber\\
&&c_6 = \frac{1}{(D+1)(D-2)} \Bigg[
\frac{(D-5)(DF_1(y^2)-(D-1)F_3(y^2))+(D-1)F_2(y^2)}{D-3}
\nonumber\\
&&\quad{}+DF_4(y^2)-(D-1)F_5(y^2)+D(A'(y^2)+2y^2A''(y^2))
+ O(x) \Bigg].
\nonumber
\end{eqnarray}
Substituting the $y^2$ expansions of the bilocal gluon condensates~(\ref{AB}),
(\ref{ABres}), (\ref{BiGe}) into~(\ref{NcGx}), (\ref{NcGcx}), we obtain
expansions consistent with~(\ref{NcG1}), (\ref{NcGc}); this provides
a strong check of our results.

\section{Heavy quarks}
\label{SecHeavy}\setcounter{equation}{0}

\subsection{Heavy quark condensates}
\label{SubCond}

If the quark mass is large, all correlators can be expressed via gluon
condensates only. This is also true for quark condensates (one--current
correlators):
\begin{equation}
Q_k=\sum_n c_{kn}(m)G_n.
\label{OPEcond}
\end{equation}
This is an expansion in $1/m$.

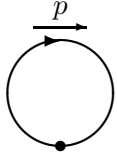
\begin{figure}[ht]
\begin{picture}(40000,7000)
\THICKLINES
\global\Xone=3983
\global\Yone=3000
\put(20000,\Yone){\circle{\Xone}}
\global\advance\Yone by -1991
\put(20000,\Yone){\circle*{400}}
\global\advance\Yone by \Xone
\drawarrow[\E\ATTIP](20000,\Yone)
\THINLINES
\global\advance\Yone by 500
\drawline\fermion[\E\REG](19000,\Yone)[2000]
\drawarrow[\E\ATTIP](\pbackx,\pbacky)
\global\advance\Yone by 500
\put(20000,\Yone){\makebox(0,0)[b]{$p$}}
\end{picture}
\caption{One-loop diagram for a heavy-quark condensate}
\label{FigCond}
\end{figure}

In the one-loop approximation (Fig.~\ref{FigCond}) quark condensates
are given by the formula
\begin{equation}
Q_k=\qq{O_k[D_\mu]}=-i\int\Dp\va{\Tr O_k[-ip_\mu-iA_\mu]S(p)},
\label{Qcond}
\end{equation}
where $A_\mu$ is given by~(\ref{Sp}). We use the \MS{} regularization:
the space dimension is $D=4-2\ep$,
\begin{equation}
\Dp\equiv\left(\frac{\mu^2e^\ga}{4\pi}\right)^\ep\frac{d^Dp}{(2\pi)^D},
\label{dp}
\end{equation}
$\mu$ is the normalization point, $\ga$ is the Euler's constant.
We substitute the quark propagator~(\ref{Sp}), (\ref{Sk}) into~(\ref{Qcond})
and average the integrand over $p$ directions in the $D$--dimensional space.
The result is expressed via the integrals
\begin{eqnarray}
&&I_n=\int\Dp\frac1{(p^2-m^2)^n}=\frac{i(-1)^n(m^2)^{2-n}}{(4\pi)^2(n-1)!}
\left(\frac{m^2}{\mu^2}\right)^{-\ep}e^{\ga\ep}\Gamma(n-2+\ep),
\nonumber\\
&&I_1=\frac{im^2}{(4\pi)^2}\left(\frac1\ep-\log\frac{m^2}{\mu^2}+1\right)
+O(\ep), \quad
I_2=\frac{i}{(4\pi)^2}\left(\frac1\ep-\log\frac{m^2}{\mu^2}\right)+O(\ep),
\nonumber\\
&&I_n=\frac{i(-1)^n}{(4\pi)^2(n-1)(n-2)(m^2)^{n-2}}+O(\ep) \quad(n>2).
\label{In}
\end{eqnarray}

Coefficients $c_{kn}$ with $d_n\le d_k$ contain ultraviolet divergences:
\begin{equation}
c^{\rm bare}_{kn}=m^{d_k-d_n}\left[\ga_{kn}\left(\frac1\ep
-\log\frac{m^2}{\mu^2}\right)+c'_{kn}\right].
\label{UV}
\end{equation}
We define the quark condensate $Q_k(\mu)$ renormalized at the point $\mu$
as the sum of the series~(\ref{OPEcond}) from which $1/\ep$ poles are
removed.

We obtain the following results~\cite{GrPin2,BrGen3}
\begin{eqnarray}
&&Q^3=-\frac1{24\pi^2}\Bigg[-6m^3(L+1)N+\frac1mG^4
-\frac1{15m^3}(G^6_1-6G^6_2)
\nonumber\\
&&\quad{}+\frac1{210m^5}
(19G^8_1+16G^8_2-20G^8_3-78G^8_4+84G^8_5-4G^8_6-18G^8_7)
+O\left(\frac1{m^7}\right)\Bigg],
\nonumber\\
&&Q^5=-\frac1{12\pi^2}\Bigg[3mLG^4-\frac1m(G^6_1-G^6_2)
\nonumber\\
&&\quad{}+\frac1{10m^3}
(3G^8_1+2G^8_2-G^8_3-9G^8_4+9G^8_5-{\textstyle\frac54}G^8_6-G^8_7)
+O\left(\frac1{m^5}\right)\Bigg],
\nonumber\\
&&Q^6=-\frac1{12\pi^2}\Bigg[LG^6_2
+\frac1{20m^2}(18G^8_5-{\textstyle\frac12}G^8_6-4G^8_7)
+O\left(\frac1{m^4}\right)\Bigg],
\nonumber\\
&&Q^7_1=-\frac1{12\pi^2}\Bigg[-3m^3(L+1)G^4+\frac1{2m}G^8_1
+O\left(\frac1{m^3}\right)\Bigg],
\nonumber\\
&&Q^7_2=\frac1{16\pi^2m}(G^8_1+G^8_2-4G^8_4)+O\left(\frac1{m^3}\right),
\nonumber\\
&&Q^7_3=\frac1{4\pi^2}\Bigg[-mLG^6_1
+\frac1{6m}(G^8_3-G^8_4+2G^8_5-{\textstyle\frac12}G^8_6)
+O\left(\frac1{m^3}\right)\Bigg],
\label{QG}\\
&&Q^7_4=-\frac1{4\pi^2}\Bigg[mLG^6_2
+\frac1{6m}(4G^8_5-{\textstyle\frac12}G^8_6-G^8_7)
+O\left(\frac1{m^3}\right)\Bigg],
\nonumber\\
&&Q^8_1=-\frac1{12\pi^2}\Bigg[3m^4(L+{\textstyle\frac32})G^4
+L(G^8_1-2G^8_3-2G^8_4)+O\left(\frac1{m^2}\right)\Bigg],
\nonumber\\
&&Q^8_2=\frac1{24\pi^2}\Bigg[12m^2(L+1)G^6_1
-L(G^8_1-G^8_2-2G^8_3+2G^8_4-4G^8_5+G^8_6)+O\left(\frac1{m^2}\right)\Bigg],
\nonumber\\
&&Q^8_3=-\frac{L}{12\pi^2}G^8_6+O\left(\frac1{m^2}\right),
\nonumber\\
&&Q^8_4=-\frac{L}{12\pi^2}G^8_7+O\left(\frac1{m^2}\right),
\nonumber\\
&&Q^8_5=\frac{L}{6\pi^2}G^8_5+O\left(\frac1{m^2}\right),
\nonumber\\
&&Q^8_6=0+O\left(\frac1{m^2}\right),
\nonumber
\end{eqnarray}
where $L=\log\frac{\mu^2}{m^2}$. As we have already mentioned,
the series~(\ref{OPEcond}) for the anomalous condensate $A$ includes
only $d=8$ gluon condensates:
\begin{equation}
A=-\frac1{32\pi^2}(G^8_1+G^8_2-4G^8_4).
\label{AG}
\end{equation}

Many of these results can be obtained using various physical
considerations~\cite{GrPin1,GrPin2} instead of the described
``brute force'' method.

\subsection{Heavy quark currents' correlators}
\label{SubCorr}

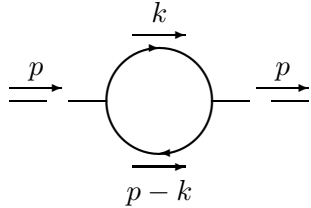
\begin{figure}[ht]
\begin{picture}(40000,8000)
\global\Xone=3983  \global\Yone=1991
\global\Xtwo=20000 \global\Ytwo=4000
\THICKLINES
\put(\Xtwo,\Ytwo){\circle{\Xone}}
\THINLINES
\global\advance\Ytwo by -\Yone
\drawarrow[\W\ATTIP](\Xtwo,\Ytwo)
\global\advance\Ytwo by -500 \global\advance\Xtwo by -1000
\drawline\fermion[\E\REG](\Xtwo,\Ytwo)[2000]
\drawarrow[\E\ATTIP](\pbackx,\pbacky)
\global\advance\Ytwo by -500 \global\advance\Xtwo by 1000
\put(\Xtwo,\Ytwo){\makebox(0,0)[t]{$p-k$}}
\global\advance\Ytwo by 1000
\global\advance\Ytwo by \Xone
\drawarrow[\E\ATTIP](\Xtwo,\Ytwo)
\global\advance\Ytwo by 500  \global\advance\Xtwo by -1000
\drawline\fermion[\E\REG](\Xtwo,\Ytwo)[2000]
\drawarrow[\E\ATTIP](\pbackx,\pbacky)
\global\advance\Ytwo by 500  \global\advance\Xtwo by 1000
\put(20000,\Ytwo){\makebox(0,0)[b]{$k$}}
\global\advance\Ytwo by -1001 \global\advance\Ytwo by -\Yone
\global\advance\Xtwo by -\Yone
\drawline\scalar[\W\REG](\Xtwo,\Ytwo)[2]
\global\advance\pbacky by 500
\drawline\fermion[\E\REG](\pbackx,\pbacky)[2000]
\drawarrow[\E\ATTIP](\pbackx,\pbacky)
\global\advance\pmidy by 500
\put(\pmidx,\pmidy){\makebox(0,0)[b]{$p$}}
\global\advance\Xtwo by \Xone
\drawline\scalar[\E\REG](\Xtwo,\Ytwo)[2]
\global\advance\pbacky by 500
\drawline\fermion[\W\REG](\pbackx,\pbacky)[2000]
\drawarrow[\E\ATTIP](\pfrontx,\pfronty)
\global\advance\pmidy by 500
\put(\pmidx,\pmidy){\makebox(0,0)[b]{$p$}}
\end{picture}
\caption{One-loop diagram for a correlator}
\label{FigGluon}
\end{figure}

Correlators of heavy quark currents can be expressed via gluon condensates:
\begin{equation}
\Pi(p)=\sum_n a_n(p^2,m)G_n.
\label{OPEheavy}
\end{equation}
The one--loop diagram (Fig.~\ref{FigGluon}) contains two propagators
of the type~(\ref{Sp}), (\ref{Sk}) (don't forget that one of them
has the vacuum momenta sink on the other side, and is given by
the formulae mirror symmetric to~(\ref{Sp}), (\ref{Sk})!).
After differentiations we obtain a formula with the integrals
\begin{equation}
\int\left(\frac{dk}{2\pi}\right)_D\frac{P(k)}{(k^2-m^2)^n((k-p)^2-m^2)^m}.
\label{Icorr}
\end{equation}
The denominators can be combined using the Feynman's formula
\begin{equation}
\frac1{a^nb^m}=\frac{\Gamma(n+m)}{\Gamma(n)\Gamma(m)}
\int\limits_0^1\frac{x^{n-1}(1-x)^{m-1}dx}{[xa+(1-x)b]^{n+m}}.
\label{Feyn}
\end{equation}
After the shift of the integration momentum $k\to k+xp$ the denominator
becomes $k^2+x(1-x)p^2-m^2$; we may average the numerator $P(k+xp)$
over $k$ directions. The integrals reduce to the form~(\ref{In}) with
$m^2\to m^2-x(1-x)p^2$. We are left with one--dimensional integrals
over the Feynman parameter $x$ of the form~\cite{NikRad}
\begin{eqnarray}
&&J_n(\xi)=\int\limits_0^1\frac{dx}{[1+\xi x(1-x)]^n}\label{Jn}\\
&&\quad{}=\frac{(2n-3)!!}{(n-1)!}\left[\left(\frac{a-1}{2a}\right)^n\sqrt{a}
\log\frac{\sqrt{a}+1}{\sqrt{a}-1}+\sum_{k=1}^{n-1}\frac{(k-1)!}{(2k-1)!!}
\left(\frac{a-1}{2a}\right)^{n-k}\right],\nonumber
\end{eqnarray}
where $\xi=-p^2/m^2$, $a=1+4/\xi$. Using these formulae, one can calculate
any gluon contribution to a heavy quark currents' correlator. The simplest
example of $d=4$ is considered in~\cite{TechRev}; contributions with
$d=6$ and $d=8$ were calculated in~\cite{NikRad}.

\section{Light quarks}
\label{SecLight}\setcounter{equation}{0}

\subsection{Limit $m\to0$ in heavy quark correlators}
\label{SubLimit}

\begin{figure}[ht]
\begin{picture}(40000,5000)
\global\Xone=1000 \global\Yone=2000
\THINLINES
\drawline\scalar[\E\REG](\Xone,\Yone)[2]
\global\Xone=\pbackx
\THICKLINES
\drawline\fermion[\N\REG](\Xone,\Yone)[2000]
\drawarrow[\N\ATBASE](\pmidx,\pmidy)
\global\advance\pfronty by -500
\put(\pfrontx,\pfronty){\makebox(0,0)[t]{$x$}}
\drawline\fermion[\E\REG](\Xone,\Yone)[10000]
\drawarrow[\W\ATBASE](\pmidx,\pmidy)
\global\advance\pmidy by -500 \global\advance\pmidx by -1000
\THINLINES
\drawline\fermion[\E\REG](\pmidx,\pmidy)[2000]
\drawarrow[\W\ATTIP](\pfrontx,\pfronty)
\global\advance\pmidy by -500
\put(\pmidx,\pmidy){\makebox(0,0)[t]{$p$}}
\global\advance\Xone by 10000
\THICKLINES
\drawline\fermion[\N\REG](\Xone,\Yone)[2000]
\drawarrow[\S\ATBASE](\pmidx,\pmidy)
\global\advance\pfronty by -500
\put(\pfrontx,\pfronty){\makebox(0,0)[t]{$0$}}
\THINLINES
\drawline\scalar[\E\REG](\Xone,\Yone)[2]
\global\Xone=\pbackx
\global\advance\Xone by 1000
\put(\Xone,\Yone){\makebox(0,0)[l]{$+$}}
\global\advance\Xone by 1500
\drawline\scalar[\E\REG](\Xone,\Yone)[2]
\global\Xone=\pbackx
\THICKLINES
\drawline\fermion[\N\REG](\Xone,\Yone)[2000]
\drawarrow[\S\ATBASE](\pmidx,\pmidy)
\global\advance\pfronty by -500
\put(\pfrontx,\pfronty){\makebox(0,0)[t]{$x$}}
\drawline\fermion[\E\REG](\Xone,\Yone)[10000]
\drawarrow[\E\ATBASE](\pmidx,\pmidy)
\global\advance\pmidy by -500 \global\advance\pmidx by -1000
\THINLINES
\drawline\fermion[\E\REG](\pmidx,\pmidy)[2000]
\drawarrow[\W\ATTIP](\pfrontx,\pfronty)
\global\advance\pmidy by -500
\put(\pmidx,\pmidy){\makebox(0,0)[t]{$p$}}
\global\advance\Xone by 10000
\THICKLINES
\drawline\fermion[\N\REG](\Xone,\Yone)[2000]
\drawarrow[\N\ATBASE](\pmidx,\pmidy)
\global\advance\pfronty by -500
\put(\pfrontx,\pfronty){\makebox(0,0)[t]{$0$}}
\THINLINES
\drawline\scalar[\E\REG](\Xone,\Yone)[2]
\end{picture}
\caption{Quark condensates' contribution to a correlator}
\label{FigQuark}
\end{figure}
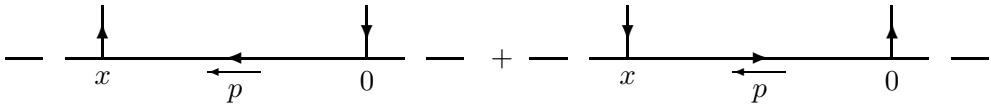

Let us consider the heavy--quark correlator~(\ref{OPEheavy})
(Fig.~\ref{FigGluon}) at $p^2\gg m^2$.
We can express it via both gluon and quark condensates:
\begin{eqnarray}
&&\Pi(p)=\Pi_G(p)+\Pi_Q(p),
\label{OPElight}\\
&&\Pi_G(p)=\sum_n a'_n(p^2,m)G_n, \quad
\Pi_Q(p)=\sum_k b_k(p^2,m)Q_k.
\nonumber
\end{eqnarray}
Here $\Pi_G(p)$ corresponds to the contribution of the region where
the virtualities of both quark and antiquark in Fig.~\ref{FigGluon}
are large ($k^2\sim p^2$), and $\Pi_Q(p)$ corresponds to the contribution
of the regions where either the quark or the antiquark has a small
virtuality ($k^2\sim m^2$). These contributions are usually depicted
as the diagrams of Fig.~\ref{FigQuark}.

The quark condensates' contribution (Fig.~\ref{FigQuark}) in the coordinate
space is
\begin{equation}
\Pi_Q(x)=\va{\q(x)\Gamma S(x,0)\Gamma q(0)}+{\rm c.c.},
\label{Qcoord}
\end{equation}
where $\Gamma$ is a $\ga$ matrix, and c.c.\ means charge conjugate.
In the momentum space
\begin{equation}
\Pi_Q(p)=\qq{(1-i\D_\al\dd_\al-\frac1{2!}\D_\al\D_\be\dd_\al\dd_\be+\cdots)
\Gamma S(p)\Gamma}+{\rm c.c.}
\label{Qmom}
\end{equation}
We substitute $S(p)$~(\ref{Sp}), (\ref{Sk}) and average over $p$ directions,
and finally reduce~(\ref{Qmom}) to the basis quark condensates following
the Section~\ref{SubQuark}. The contribution of $A$ is omitted because
this condensate is a gluon one. The quark contributions with $d=7$, $d=8$
were obtained in~\cite{GrPin1,BrGen3}.

Having obtained $\Pi_Q(p)$, we can find $\Pi_G(p)$ from the heavy-quark
correlator $\Pi(p)$ using the expansions~(\ref{QG}):
\begin{eqnarray}
&&\Pi_G(p)=\sum_n a'_n(p^2,m)G_n=\Pi(p)|_G-\Pi_Q(p)|_G
\nonumber\\
&&\quad{}=\sum_n a_n(p^2,m)G_n-\sum_{k,n}b_k(p^2,m)c_{kn}(m)G_n.
\label{PiG}
\end{eqnarray}
The coefficients $a_n(p^2,m)$ have singularities at $m\to0$ arising from
the regions $k^2\sim m^2$. These singularities have to cancel in~(\ref{PiG})
giving $a'_n(p^2,m)$ finite at $m\to0$. Using the results of~\cite{NikRad},
the gluon contribution with $d=8$ was obtained in~\cite{GrPin2,BrGen3}.

\subsection{Minimal subtraction of mass singularities}
\label{SubMin}

The method of the previous Section is good when the heavy quark correlator
is already known. But when we want to calculate a light quark correlator
from scratch, there should be a simper way than to solve a more difficult
heavy quark problem first. Such a method was proposed in~\cite{BrGen2},
and used in~\cite{BrGen3} for $d=8$ calculations.

Now we want to go to the limit $m\to0$ before $D\to4$. The difference
of a renormalized quark condensate and a bare one is (see~(\ref{UV}))
\begin{equation}
Q_k-Q^{\rm bare}_k=-\frac1\ep\sum_{d_n\le d_k}m^{d_k-d_n}\ga_{kn}G_n
\to-\frac1\ep\sum_{d_n=d_k}\ga_{kn}G_n.
\label{RenBare}
\end{equation}
At $m=0$ all loop integrals for $Q^{\rm bare}_k$ vanish because they
contain no scale (ultraviolet and infrared divergences cancel each other).
Therefore we obtain for the gluon contribution to a correlator
(see~(\ref{PiG}))
\begin{equation}
\Pi_G(p)=\Pi(p)|_G+\frac1\ep\sum_{d_n=d_k}b_k(p)\ga_{kn}G_n.
\label{MSMS}
\end{equation}
Here $\Pi(p)|_G$ is calculated in the \MS{} scheme with $m=0$,
and $\ga_{kn}$ are mixing coefficients of quark condensates $Q_k$
with gluon condensates $G_n$ of the same dimension (coefficients at $L$
in~(\ref{QG})). Omitting $1/\ep$ poles we finally obtain
\begin{equation}
\Pi_G(p)=\Pi(p)|_G+\sum_{d_n=d_k}\frac{db_k}{d\ep}\ga_{kn}G_n.
\label{MSMS2}
\end{equation}
The quark condensates' coefficient functions $b_k$ should be calculated
at $m=0$ up to linear terms in $\ep$ (using $D$--dimensional tensor
and $\ga$ matrix algebra, and in particular $D$--dimensional averaging,
see Section~\ref{SubAver}).

{\bf Acknowledgements.} I am grateful to Yu.~F.~Pinelis for the collaboration
in writing the papers~\cite{GrPin1,GrPin2}, to D.~J.~Broadhurst
for useful discussions of the works~[4--7], 
and to S.~V.~Mikhailov for fruitful discussions of nonlocal condensates.
Various stages of this work were supported by grants from International
Science Foundation, Russian Foundation of Fundamental Research, Royal
Society, and PPARC.


\newpage

\begin{thebibliography}{99}

\bibitem{TechRev}                                                
V.~A.~Novikov, M.~A.~Shifman, A.~I.~Vainshtein, V.~I.~Zakharov.
Fortschr.\ Phys.\ 32 (1984) 585.

\bibitem{NikRad1}                                                
S.~N.~Nikolaev, A.~V.~Radyushkin.
\PLB110 (1982) 476; (E) B116 (1982) 469; \NPB213 (1983) 285.

\bibitem{NikRad}                                                 
S.~N.~Nikolaev, A.~V.~Radyushkin.
\PLB124 (1983) 243; Sov.\ J.\ Nucl.\ Phys.\ 39 (1984) 91.

\bibitem{BrGen1}                                                 
S.~C.~Generalis, D.~J.~Broadhurst.
\PLB139 (1984) 85.

\bibitem{BrGen2}                                                 
D.~J.~Broadhurst, S.~C.~Generalis.
\PLB142 (1984) 75.

\bibitem{BrGen3}                                                 
D.~J.~Broadhurst, S.~C.~Generalis.
\PLB165 (1985) 175.

\bibitem{Gen}                                                    
S.~C.~Generalis. Open University thesis OUT-4102-13 (1984);
J.\ Phys.\ G15 (1990) L225; G16 (1990) 367, 785, L117.

\bibitem{GrPin1}                                                 
A.~G.~Grozin, Yu.~F.~Pinelis.
\PLB166 (1986) 429.

\bibitem{GrPin2}                                                 
A.~G.~Grozin, Yu.~F.~Pinelis.
Zeit.\ Phys.\ C33 (1987) 419.

\bibitem{Old}                                                    
V.~N.~Baier, Yu.~F.~Pinelis. Preprint INP 81-141, Novosibirsk (1981);
D.~Gromes. \PLB115 (1982) 482;
E.~V.~Shuryak. \NPB203 (1982) 116.

\bibitem{MikhRad}                                                
S.~V.~Mikhailov, A.~V.~Radyushkin. JETP Lett.\ 43 (1986) 712;
Sov.\ J.\ Nucl.\ Phys.\ 49 (1989) 494;
Phys.\ Rev.\ D45 (1992) 1754.

\bibitem{Appl}                                                   
S.~V.~Mikhailov, A.~V.~Radyushkin. Sov.\ J.\ Nucl.\ Phys.\ 52 (1990) 697;
A.~V.~Radyushkin. \PLB271 (1991) 218;
A.~P.~Bakulev, A.~V.~Radyushkin. \PLB271 (1991) 223;
A.~G.~Grozin, O.~I.~Yakovlev. \PLB285 (1992) 254; B291 (1992) 441;
A.~P.~Bakulev, S.~V.~Mikhailov. Preprint JINR E2--94--208, Dubna (1994);
hep-ph/9406216.

\bibitem{Simonov}                                                
H.~G.~Dosh, Yu.~A.~Simonov. \PLB205 (1988) 338;
Yu.~A.~Simonov. \NPB307 (1988) 512; B324 (1989) 67;
Sov.\ J.\ Nucl.\ Phys.\ 50 (1989) 134; 54 (1991) 115.

\bibitem{Mikh}                                                   
S.~V.~Mikhailov. Phys.\ Atom.\ Nucl.\ 56 (1993) 650.

\end{thebibliography}
\end{document}